\documentclass{aa}  
\usepackage{float}

\usepackage{multirow}
\usepackage{booktabs}
\usepackage{graphicx}
\usepackage{txfonts}
\usepackage{lipsum}
\usepackage{subcaption}

\usepackage{lscape} 

\usepackage{placeins}

\usepackage[normalem]{ulem}

\usepackage{siunitx}
\usepackage{xcolor}

\begin{document}

   \title{Multivariate time series transformer embeddings for light curves of periodic variable stars}

   \author{G. Chiong\inst{1}
        \and I. Becker\inst{1}
        \and P. Protopapas\inst{1}
        }

   \institute{John A. Paulson School of Engineering and Applied Science, Harvard University, Cambridge, MA, 02138\\
             \email{pavlos@seas.harvard.edu}\\ }

   \date{Received June 13, 2025}

  \abstract
   {Astronomical surveys produce time series data by observing stellar objects across multiple photometry bands. Foundational transformer-based models, such as \texttt{Astromer}, encode each time series as a sequence of embeddings to classify sources. However, such models operate independently on each band and therefore do not use information across wavelengths or filters.}
   {We extend the single-band \texttt{Astromer} framework by introducing a fusion layer that combines single photometry band observations into a unified sequence representation, thus enabling multiband analysis for downstream tasks, namely periodic variable star classification. The challenge in adapting the encoder for multiband data lies in coordinating information across bands observed at asynchronous times.}
   {We pre-trained various multiband neural network models on 600 000 high signal-to-noise light curves from the Massive Compact Halo Object (\texttt{MACHO}) survey and fine-tuned them using labeled data from the \texttt{Alcock} catalog (derived from \texttt{MACHO}) and from the Asteroid Terrestrial-impact Last Alert System survey.}
   {Our results show that both proposed multiband architectures outperform the single-band models by approximately ten percentage points in the F1-score. Jointly pre-trained multiband encoders further improve the performance compared to a collection of independently pre-trained single-band encoders, reaching an improvement of $23$ percentage points in the F1-score, even when data are scarce. Furthermore, we find minimal differences in multiband performance between sampling bands asynchronously and sampling all bands on the same time steps, as most models remain within a three percentage point difference in F1-score on the full dataset.}
   {These results demonstrate a trade-off in the training speed and classification accuracy between single-band and multiband encoders, with multiband models improving the F1-score performance by ten percentage points at the cost of increased pre-training time. Our results show the great potential of transformer-based multiband neural network architectures for classification tasks regarding future large-scale time-domain surveys with a greater variety of variable stars.}

   \keywords{representation learning --
                multivariate light curves --
                foundational models
               }

   \maketitle
   \nolinenumbers

\section{Introduction}

Variable stars are astronomical objects whose brightness changes over time due to intrinsic phenomena (e.g., pulsations or eruptions) or extrinsic factors (e.g., eclipses in binary systems). These flux variations encode information about an object's structure, composition, and state. A light curve is a record of this variation of the light received over a period of time from a variable star. Analyzing the brightness variations of stellar objects over time can yield useful statistical properties for tasks such as the study of physical processes within objects or classifying them into distinct categories or types (see \citealp{catelan_2015} for a review).

The systematic analysis and classification of variable stars play an important role in many areas of astronomy. Variable objects are used to probe stellar evolution, characterize stellar populations, and calibrate distance indicators such as Cepheids and RR Lyrae stars \citep{catelan_2015}. At larger scales, time-domain measurements contribute to studies of galactic structure and evolution \citep{gaia_2016}, and provide essential inputs for cosmological analyses based on standard candles and transient populations \citep{villar_2020}. As modern surveys increasingly observe the sky repeatedly and across multiple photometric bands, the ability to robustly model and classify astronomical time series data has become a central challenge in observational astronomy.

Traditionally, light curve observations were recorded in a single band to maximize temporal coverage. This has resulted in methods of light curve analysis that predominantly rely on manually engineered statistical and time-domain features specific to a single band \citep{richards_2011, kim_2014, sanchez-saez_2019, forster_2021}. These techniques, however, often require substantial domain expertise, and feature engineering is a resource-intensive process \citep{graham_2014, nun_2017, sanchez-saez_2021}.

The advent of large-scale astronomical surveys, such as Gaia \citep{gaia_2016}, the Asteroid Terrestrial-impact Last Alert System \citep[\texttt{ATLAS};][]{tonry_2018}, the Panoramic Survey Telescope and Rapid Response System \citep{chambers_2019}, the Zwicky Transient Facility \citep{bellm_2019}, and the upcoming Legacy Survey of Space and Time \citep[LSST;][]{ivezic_2019} by the Vera C. Rubin Observatory, have increased the availability of multiband light curve datasets. These surveys employ a variety of different filters, and they capture complementary information since they are designed to map different portions of the electromagnetic spectrum \citep[e.g., ][]{1996AJ....111.1748F}.

Multiband light curve analysis is complicated by the inherent asynchronous nature of observations across different bands \citep{Bianco_2022}. Measurement uncertainties further exacerbate these issues, as the different bands usually show different behaviors. Techniques designed for multiband data must either explicitly account for band-specific timestamps or adopt strategies such as interpolation or learned embeddings to integrate information across bands at any given time step.

Various time series modeling and data processing techniques have been developed to address the challenges posed by irregular and asynchronous multiband sampling. These can be broadly categorized into three strategies:

\begin{itemize}
\item Imputation and interpolation: These methods fill in missing or irregularly sampled values to produce a uniform sequence. Examples include random imputation between past and future observations \citep{charnock_2017}, linear interpolation \citep{muthukrishna_2019}, and probabilistic modeling using Gaussian processes \citep{boone_2019, villar_2020, jamal_2020}.

\item Explicit masking: To preserve the asynchronous nature of the data without introducing artificial values, some models use binary masking schemes that indicate whether an observation is present in a given band at each time step \citep{moller_2019, pimentel_2023, cabrera-vives_2024}.

\item 
Learned representations: Other approaches use deep learning models, including Recurrent neural network (RNN)-based architectures, to learn unified embeddings directly from raw multiband light curves, accommodating irregular and asynchronous sampling \citep{donoso-olivia_2021, becker_2025}.
\end{itemize}

Among these approaches, deep learning methods have gained particular traction for their ability to learn informative representations directly from light curves without relying on imputation or manually engineered features. Recurrent neural networks, including Long short-term memory (LSTM) and Gated recurrent unit, have been widely used for astronomical time series modeling and classification tasks \citep{hochreiter_1997, cho_2014}.

Early efforts on light curve modeling and variable-star classification employed RNN-based architectures \citep{naul_2018, becker_2020, moller_2019, donoso-olivia_2021, jamal_2020}; however, RNNs face well-known limitations in capturing long-range dependences and in training parallelism \citep{pascanu_2013}. The transformer architecture \citep{vaswani2017attention} replaces recurrence with self-attention, computing pairwise interactions between all elements of the input sequence in parallel, which has motivated its recent adoption for irregular astronomical time series.

To address such complexities and leverage the strengths of transformer architectures, recent research has turned to foundational models \citep{bommasani_2022}. These models learn general-purpose representations that can be fine-tuned for a wide range of downstream tasks, such as classification or regression. This family of models has the ability to model diverse sequence types without requiring labeled data, making them ideal in data-rich but annotation-sparse fields such as astronomy.

Training foundational models from scratch requires significant computational resources and large-scale data. However, once pre-trained, these models can be fine-tuned on smaller, domain-specific datasets, reducing training cost and time for downstream applications.

Transformer-based foundational models, inspired by architectures such as Bidirectional encoder representations from transformers \citep[BERT;][]{devlin_2019} and Generative pre-trained transformers \citep{gpt}, have shown that large-scale pre-training can produce transferable sequence representations across domains. Much recent work has focused on building transformer-based models as a tool for light curve classification. \cite{morvan_2022} used a transformer model to remove noise and outliers in time series light curves, and \cite{pimentel_2023} proposed an attention-based model to classify multiband light curves of different transient event types. \cite{cadiz-leyton_2024} investigated combining uncertainty estimation techniques with a pre-trained astronomical time series transformer to classify variable objects. More recently, \cite{cabrera-vives_2024} proposed ATAT, a transformer-based architecture that jointly models multiband light curves and tabular metadata for variable-source classification.

\texttt{Astromer 1}~(hereinafter~\texttt{Astromer})~\citep{donoso-olivia_2023} is a transformer-based foundational model. It operates directly on sequences of continuous magnitude measurements and their associated observation times for classification using single–band photometry.

Although \texttt{Astromer} has been shown to perform well at sequence modeling, its architecture is not designed to process asynchronously sampled multiband data. In this work, we introduce \texttt{Multiband Astromer}, an extension of \texttt{Astromer} that is able to handle the analysis of multiband light curves for periodic variable stars, and we evaluate the classification performance of \texttt{Multiband Astromer} using the constructed datasets from the Massive Compact Halo Object \citep[\texttt{MACHO};][]{alcock_1993} survey and \texttt{ATLAS} survey \citep{heinze_2018}.

The proposed architecture comprises three main components. First, each photometric band is encoded using a transformer-based model. Second, the resulting embeddings are combined by a dedicated layer, which produces a unified representation that leverages cross-band information. Third, the encoded representations are used for downstream classification tasks, similar to the original \texttt{Astromer} model.

In summary, the main contribution of this work is to present the results of characterizing multiband light curves by proposing a novel transformer-based framework that fuses information across photometric bands. The remainder of this paper is organized as follows:

Section \ref{sec:astromer-intro} introduces the original \texttt{Astromer} model and outlines its architectural design for single-band light curves. Section \ref{sec:mb-astromer} then presents the proposed \texttt{Multiband Astromer} framework, including a description of the variations explored in pre-training strategies, embedding fusion architectures, and time series sampling techniques. Section \ref{sec:data} follows with a description of the datasets and experimental methodology, encompassing pre-training, fine-tuning, and evaluation metrics. The results and analyses are presented in Section \ref{sec:results}, where the performance of the multiband models is compared against single-band baselines and the influence of synchronous versus asynchronous observation sampling is examined. Finally, Section \ref{sec:conclusion} concludes the paper with a summary of the key findings and an outline of the potential directions for future applications.

\section{The original \texttt{Astromer} architecture} \label{sec:astromer-intro}

\texttt{Astromer} \citep{donoso-olivia_2023} is a foundational transformer-based model developed to encode single-band astronomical light curves into dense vector representations. Drawing inspiration from BERT \citep{devlin_2019}, \texttt{Astromer} adapts the transformer architecture to continuous time series light curve data.

Prior to describing our extension of \texttt{Astromer} into the multiband paradigm in Sect.~\ref{sec:mb-astromer}, we first introduce the basic ideas of its architecture for the single-band scenario. We briefly summarize the components most relevant to our multiband extension.

\texttt{Astromer} is pre-trained on 1.5 million R-band light curves obtained from the \texttt{MACHO} survey \citep{alcock_1993} using a masked reconstruction strategy, where the model predicts the values of randomly masked observations. This is later fine-tuned on labeled datasets to solve downstream tasks such as classification, which updates the encoder to better represent a specific target dataset.

To introduce the notation used later in this paper to describe our input data, we denote each light curve in band $i$ by $ Z_i^{(k)} \in \mathbb{R}^{m \times 2} $, where $m$ is the number of temporal observations and the two columns correspond to observation time and magnitude. This sequence is formalized as:

\begin{equation}
    Z_i^{(k)} = \left\{(t_{i,0}^{(k)}, x_{i,0}^{(k)}), (t_{i,1}^{(k)}, x_{i,1}^{(k)}), \dots, (t_{i,j}^{(k)}, x_{i,j}^{(k)}), \ldots ,(t_{i,m-1}^{(k)}, x_{i,m-1}^{(k)})\right\}.
\end{equation}
Here, each $ t_{i,j}^{(k)} $ denotes the modified Julian date (MJD) corresponding to the $ j $-th observation of light curve $ k $ under filter $ i $, and $ x_{i,j}^{(k)} $ represents the respective observed magnitude. As \texttt{Astromer} is designed only for single-band data, we omit the band $i$ notation here in Sect.~\ref{sec:astromer-intro} for simplicity.

Typically, a survey's cadence is nonuniform, making the number of observations $m$ different for each object. To maintain a model of manageable size, \texttt{Astromer} randomly samples contiguous fixed-length sequences of $L = 200$ observations for the chosen band. Sequences with $ L < 200$  are padded with zeros to complete L elements.\footnote{These values were selected to match the original \texttt{Astromer} implementation \citep{donoso-olivia_2023}, where they were fine-tuned.}

\subsection{Model architecture}
\begin{figure}
    \centering
    \includegraphics[scale=.5]{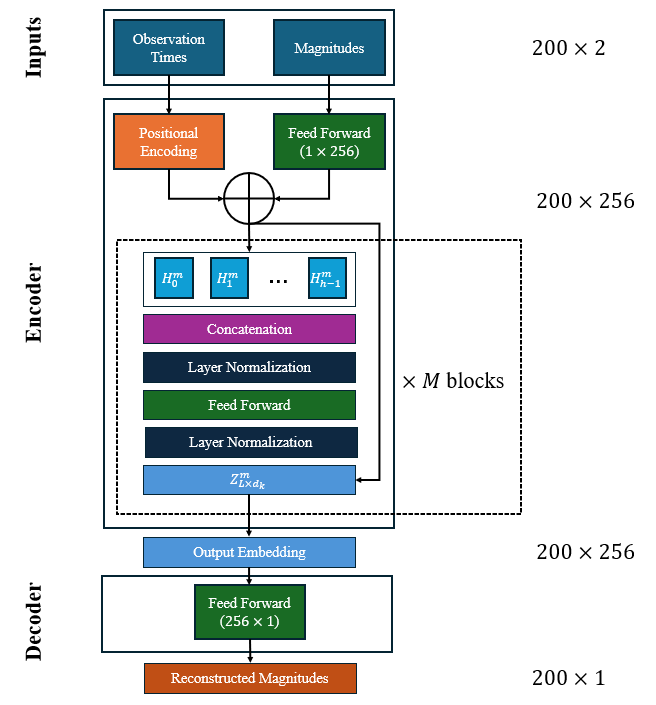}
    \caption{
    \texttt{Astromer} architecture for the masked reconstruction task. The dimension of the input of each block is shown on the right. The default configuration uses \(h = 4\) attention heads for each of the \(M = 2\) encoder layer blocks. The output of the first block is the input to the second block and so forth. The encoder layer’s output embedding is derived from the final block. The decoder can be modified to suit the desired downstream task.}
    \label{fig:astromer-architecture}
\end{figure}

\texttt{Astromer} is designed as an encoder composed of two self-attention blocks. Each block contains four attention heads, with each head operating in a subspace of dimension $d_h = 64$$^1$; the final embedding dimension is the concatenation of all the 256 attention heads$^1$. Figure \ref{fig:astromer-architecture} illustrates the two-layer encoder architecture, showing how temporal and magnitude inputs are transformed into contextual embeddings through multi-head self-attention, in which several attention heads operate in parallel to compute pairwise relationships between observations in the sequence.

The \texttt{Astromer} encoder represents each light curve as a sequence of 256-dimensional embeddings by combining projected magnitudes with sinusoidal encodings of observation times. The sequence is processed using multi-head self-attention with standard scaled dot-product attention and masking during pre-training. The encoder outputs a fixed-length sequence of contextual embeddings that capture temporal and photometric dependences within the light curve. Further implementation details are provided in Appendix~\ref{appendix:astromer}.

\subsection{Training strategy} \label{subsec:astromer-training}
The original \texttt{Astromer} \citep{donoso-olivia_2023} is initially pre-trained and used to reconstruct light curves. As in the original implementation, the adopted masking strategy is designed to prevent the network from learning an identity mapping. The masking protocol is as follows\footnote{These values were also selected to match the original \texttt{Astromer} implementation \citep{donoso-olivia_2023}, where they were fine-tuned.}: $50\%$ of the observations in the sequence are randomly masked. Among these masked positions, $20\%$ are substituted with their corresponding original observations. Additionally, $10\%$ of the masked positions are replaced with random values.

The reconstruction loss is computed exclusively over the masked subset,
\begin{equation}
    \mathcal{L} = \sqrt{\frac{1}{B} \sum_{b=0}^{B-1} \sum_{l=0}^{L-1} \mu_{b,l} \,\bigl(x_{b,l} - \hat{x}_{b,l}\bigr)^2},\label{eq:rmse-loss}
\end{equation}
where $B$ is training batch size, $b$ is the index of a light curve within the training batch, $l$ the time step index within the light curve, $\mu_{b,l}$ is a binary indicator specifying whether the observation $ x_{b,l} $ has been masked, and $ \hat{x}_{b,l} $ is the corresponding model prediction. We pre-train the model on 1.5 million unlabeled light curves from the \texttt{MACHO} survey using the Adam optimizer \citep{kingma_2014} with a learning rate of $10^{-3}$ and early stopping based on the root-mean square error (RMSE) of the validation dataset.

Fine-tuning is subsequently performed on a task-specific dataset, which usually contains many fewer samples. This allows the model to adapt its weights to the new data. The loss in fine-tuning is calculated in the same way as pre-training (see Eq. \ref{eq:rmse-loss}).

\subsection{\texttt{Astromer} performance and applications}
The effectiveness of the learned embeddings is evaluated in classification tasks through two distinct architectures by swapping the feed forward decoder layer used for masked reconstruction pre-training accordingly: an LSTM and attention mechanism (ATT) classifier (LSTM + ATT) that processes the sequence of attention vectors, and a multilayer perceptron classifier (MLP + ATT) that operates on averaged attention embeddings along the time dimension. Classification performance is evaluated using the F1-score, which is the harmonic mean of precision and recall, providing a single metric that balances both. It is defined as
\begin{equation}\label{eq:f1-score}
    \text{F1-score} = 2 \times \frac{P \times R}{P + R},
\end{equation}
where \( P = \frac{TP}{TP + FP} \) is the precision, and \( R = \frac{TP}{TP + FN} \) is the recall. Here, \( TP \), \( FP \), and \( FN \) denote the number of true positives, false positives, and false negatives, respectively. The F$_1$ score ranges from 0 to 1, with higher values indicating better overall classification performance.
The classification loss is the cross-entropy, and it is shown in Eq. \ref{eq:cross-entropy}:
\begin{equation}\label{eq:cross-entropy}
    CE = -\sum_{i=1}^{K}{y_i\log{\hat{y}_i}},
\end{equation}
where $K$ is the number of classes, $y$ denotes the one-hot encoded ground-truth label, and $\hat{y}_i$ is the predicted probability for class $i$ produced by the classifier.

This completes our overview of the original \texttt{Astromer} model \citep{donoso-olivia_2023}. In the following section, we extend this framework to the multiband models.

\section{\texttt{Multiband Astromer}} \label{sec:mb-astromer}

We extend the \texttt{Astromer} model to the multiband regime by introducing two architectural frameworks --- the simple multiband architecture (\texttt{SMA}) and the full multiband architecture (\texttt{FMA}) --- to generate unified embeddings from multiband light curves for downstream classification.

\subsection{Overview of model architectures}

The \texttt{SMA} framework consists of multiple independently pre-trained \texttt{Astromer} encoders, each tailored to process light curves from a distinct band. In this design, the encoders are trained separately, with no information shared during the pre-training phase. After encoding, the embeddings generated by each band-specific encoder are combined using a late fusion embedding mixing (LFEM).\footnote{The LFEM refers to a general design pattern in which band-specific encoders are fused only after encoding. This allows for flexibility in implementing a variety of mixing strategies within the model.} This approach defers the integration of the band-specific information until the downstream task.

The \texttt{FMA} framework shares the same architectural components, a set of single-band encoders and a fusion mechanism, but differs from \texttt{SMA} in training. Both architectures use the same combination mechanism (LFEM) to combine band-specific embeddings; however, in \texttt{SMA}, this fusion occurs only after the encoders are pre-trained independently. In contrast, \texttt{FMA} uses the fusion output during pre-training to reconstruct all bands jointly. Gradients are backpropagated through the fusion layer to all encoders, enabling coordinated updates and allowing the model to learn cross-band relationships from the beginning.

In this sense, the key difference between \texttt{SMA} and \texttt{FMA} lies not in architecture, but in their training strategy. \texttt{SMA} learns band-specific features independently and integrates them only during fine-tuning. \texttt{FMA}, by contrast, trains encoders jointly via a shared reconstruction objective, enabling it to model cross-band correlations earlier in the learning process.

At an architectural level, the design of \texttt{FMA} is closely related to earlier multiband classification approaches that concatenate per-band representations prior to downstream analysis, such as the hybrid architecture described by \citet{jamal_2020}. In that work, the equivalent of our mixing layer is implemented via feature concatenation, and all network components are trained jointly.
Our contribution differs in that we explicitly decouple architectural design from training strategy, and systematically study the impact of independent versus joint pre-training of band-specific encoders under varying data availability.

\begin{figure}
    \centering
    \includegraphics[scale=.35]{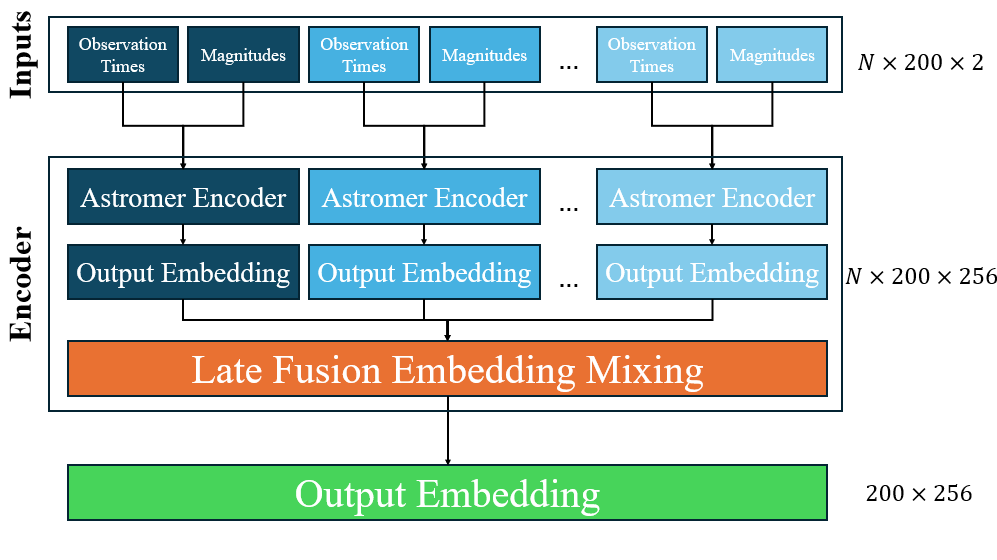}
    \caption{
    \texttt{Multiband Astromer} encoder architecture for one light curve with matrix dimensions. The encoder ingests $N$ input bands, each consisting of $L=200$ paired observation times and magnitudes. Each band is processed by one distinct \texttt{Astromer} encoder, which projects the input data into a high-dimensional embedding. Following the LFEM design pattern, the fusion stage can be instantiated using various mixing strategies (e.g., AVG, LF, and CXA), as described in Sect.~\ref{subsec:emb-mix}.
    }
    \label{fig:multiband-architecture}
\end{figure}

A comparative analysis of the \texttt{SMA} and \texttt{FMA} models is presented in Sect.~\ref{subsec:arch-compare}. Figure \ref{fig:multiband-architecture} shows an overview of the multiband architecture.

During both the pre-training and fine-tuning phases, the model combines the information to reconstruct the masked observations from each input light curve. Specifically, an encoder $\Phi_i$ for band $i$ computes an output
\begin{equation}
    z_i^{(k)} = \Phi_i(Z_i^{(k)}).
\end{equation}

Each band-specific encoder $\Phi_i$ maps its input $Z_i^{(k)} \in R^{L\times2}$ to a sequence embedding $z_i^{(k)} \in \mathbb{R}^{L \times d}$, where $L$ refers to the number of observations and $d$ to the dimensionality of the encodings. These embeddings are then combined using a mixing function $f$ to produce a unified embedding:
\begin{equation}
    y = f\left(z_0^{(k)}, z_1^{(k)}, \dots, z_{N-1}^{(k)}\right).
\end{equation}
The decoder $g$ then reconstructs masked observations from this shared representation, and the reconstruction losses are averaged across bands:
\begin{equation}
    \mathcal{L} = \frac{1}{N} \sum_{i=0}^{N-1} \mathcal{L}_i,
\end{equation}
where $\mathcal{L}_i$ refers to the masked reconstruction loss of band $i$ in Eq. \ref{eq:rmse-loss}.

This aggregated loss is backpropagated through the $f$ layer and each encoder, thereby updating the model parameters in an end-to-end manner. Given that the encoder's output embedding dimension is the same as the \texttt{Astromer} model's, we reused the same head structure for the downstream classification task.

\subsection{Embedding mixing strategies}\label{subsec:emb-mix}

The choice of mixing strategy is critical to the model’s ability to integrate multiband information effectively. We explore several approaches, focusing on the trade-off between computational cost and representational power. The objective of the mixing strategy is to generate a unified embedding that effectively captures cross-band information from an ensemble of single-band \texttt{Astromer} models, with the aim of improving performance in downstream tasks. We present our findings in a two-band setting for simplicity, illustrating using the \texttt{MACHO} dataset's red and blue bands \citep{alcock_1993}, and the \texttt{ATLAS} dataset's orange and cyan bands \citep{tonry_2018}. This architecture can be extended to more bands without major changes.

In this work, several mixing methods are explored to balance the cost of inference versus the model's capacity to learn cross-band dependences. We use the following short-hand notation to describe the mixing strategies:
\begin{itemize}
    \item AVG: Element-wise averaging
    \item LF: Learnable fusion
    \item CXA: Cross-attention
    \item MHA: Cross-attention with multi-head attention
    \item LSTM: Cross-attention with LSTM
    \item TF: Transformer
\end{itemize}
Furthermore, the mixing strategy notation is prefixed with the architecture (i.e., \texttt{FMA} or \texttt{SMA}) separated by a hyphen to specify a model used in the experiments. As an example, \texttt{FMA-AVG} refers to the Full Multiband Architecture with the Element-wise Averaging mixing strategy.

\paragraph{Element-wise averaging}
A natural baseline is element-wise averaging, where embeddings $z_i^{(k)}$ from $N$ bands are aggregated as
\begin{equation}
    z_{\text{avg}} = \frac{1}{N} \sum_{i=0}^{N-1} z_i^{(k)}.
\end{equation}
While the AVG mixing layer does not have any trainable parameters (and is therefore the most computationally efficient strategy), it presumes equal importance of each band, does not explicitly account for interactions between bands, and there is no guarantee that the embedding dimensions contain the same information. Furthermore, it uniformly carries out embedding averaging even when observation times are asynchronous.

\paragraph{Learnable fusion}
The learnable fusion mechanism is based on a multi-layer perceptron (MLP). The embeddings are concatenated along the embedding dimension and processed through a three-layer feedforward neural network with $N \times d$, $1024$, and $512$ units, respectively, using the $\operatorname{ReLU}$ activation function. Dropout \citep{srivastava_2014} and layer normalization \citep{ba_2016} are incorporated for regularization.

\paragraph{Cross-attention and multi-head attention}
An alternative mixing strategy leverages cross-attention to directly process cross-band interactions. Cross-attention \citep{lin_2021} is a mechanism in transformer models that computes attention scores between a query $Q$ sequence and a separate key $K$ and value $V$ sequence, allowing information transfer between different data sources. Specifically, $Q$ is projected from one single-band encoder, while $K$ and $V$ are projected from another single-band encoder.

Given embeddings $z_R$ and $z_B$ from the $R$ and $B$ bands, cross-attention is computed by treating $z_R$ as the query and $z_B$ as both keys and values, and vice versa:
\begin{equation}
    z_{R \to B} = \operatorname{softmax}\!\left(\frac{Q_R K_B^\top}{\sqrt{d_k}}\right) V_B, \quad
    z_{B \to R} = \operatorname{softmax}\!\left(\frac{Q_B K_R^\top}{\sqrt{d_k}}\right) V_R.
\end{equation}

Note that the computational complexity of this mixing strategy architecture scales quadratically with the number of bands. This is because the cross-attention mechanism computes each band's embedding against all other band embeddings.

In our CXA architecture, one block is composed of four heads, each with $128$ units, and the embeddings are then concatenated,
\begin{equation}
    z_{\text{concat}} = z_{R \to B} \mathbin\Vert z_{B \to R}.
\end{equation}
A fully connected layer is used to combine the information and project back into the $L \times d$ space. 
Our work further explores an extension of this idea in which a fully connected layer combines the cross-embeddings, after which a multi-head attention (MHA) layer further refines the representation using two heads, each with 64 units, before projecting it back into the $L \times d$ space.

\paragraph{LSTM}
Alternatively, the concatenated embeddings from the CXA mixing strategy may be processed by a bidirectional LSTM:
\begin{equation}
    z_{\text{LSTM}} = \operatorname{BiLSTM}_{512}(z_{\text{concat}}),
\end{equation}
where each directional LSTM consists of $256$ units, before being projected back into the output of dimension of $L \times d$. In our study, this is referred to as our LSTM architecture.

\paragraph{Transformer}
The last mixing technique involves concatenating the individual band embeddings and transforming the result with an additional transformer block. This block employs a self-attention layer with four heads, each of dimensionality $d_h = 128$, with residual connections and layer normalization preceding a final projection back to a $(L \times d)$-dimensional space.

\subsection{Comparative analysis of architectures}\label{subsec:arch-compare}

The \texttt{SMA} framework offers the advantage of modularity, as each band-specific encoder can be pre-trained independently by different sources. This facilitates collaboration and open-source contributions while reducing the overall training time of any single entity, as different parties can share the effort to pre-train the required single-band models. Furthermore, the sharing of pre-trained models brings environmental benefits by reducing computing carbon footprint, as pre-training only needs to be executed once. However, when coupled with basic mixing techniques, such as AVG, its capacity to capture cross-band dependences may be limited, particularly as the encoders are pre-trained without any coordination in the output embedding space.

In contrast, \texttt{FMA} incorporates joint pre-training, which promotes compatible representations between the band encoders and allows direct analysis of the cross-band relationships. However, this increased representation power comes at the cost of training time and computational resources required; particularly with the quadratic scaling of attention operations with respect to the number of bands when using cross-attention mechanisms.

Compared to the AVG approach, the LF strategy aims to improve on the performance of the baseline by concatenating band-specific embeddings and processing them through an MLP. This approach enables the model to employ band-dependent weights, thereby capturing a richer set of cross-band dependences. Nonetheless, the additional parameters incur extra computational cost compared to the non-parametric averaging method (i.e., $2.36$ million mixing layer parameters for LF and $0$ mixing layer parameters for AVG from Tab.~\ref{tab:mixing-parameters}).

The use of cross-attention offers a more explicit mechanism for aligning features from different bands by computing attention scores between them. The CXA strategy facilitates direct information exchange between the embeddings, enhancing the model’s ability to capture complementary characteristics. When extended to the MHA strategy, an additional multi-head attention module further refines the cross-attention vectors. Despite their representational advantages, both CXA and MHA are associated with a quadratic increase in computational complexity of the mixing layer as the number of bands increases.

Rather than a MHA module, a bidirectional LSTM can be applied to the concatenated outputs from the cross-attention blocks. The LSTM strategy is designed to focus on capturing sequential dependences. However, the sequential nature of LSTM may limit parallelization relative to purely attention-based approaches.

Finally, the TF mixing strategy employs an additional transformer block on the concatenated embeddings, utilizing self-attention to achieve a global integration of band-specific information. Similar to other parameterized training strategies, this method remains sensitive to the choice of hyperparameters (such as the number of heads and key dimensionality) and shares the computational overhead of attention-based methods. We present the number of model parameters for each mixing strategy in Tab.~\ref{tab:mixing-parameters}. For comparison, the rest of the model components (including encoder and decoder, excluding mixing layer) in our study combine for 2.2 million parameters in total.

\begin{table}
\centering
\caption{Number of parameters in the mixing layer for each strategy.}              
\label{tab:mixing-parameters}
\begin{tabular}{lc} 
\hline \hline
Mixing Strategy         & Mixing Layer Params.  \\
 & (millions) \\\hline
AVG   &  0              \\
LF    &  \num{2.36}  \\
CXA   &  \num{4.85}  \\
MHA   &  \num{2.43}  \\
LSTM  &  \num{2.82}  \\
TF    &  \num{2.23}  \\ \hline
\end{tabular}
\tablefoot{Excludes encoder and decoder components.}
\end{table}

\subsection{Synchronous and asynchronous sampling}

\begin{figure}
    \centering
    \includegraphics[scale=.45]{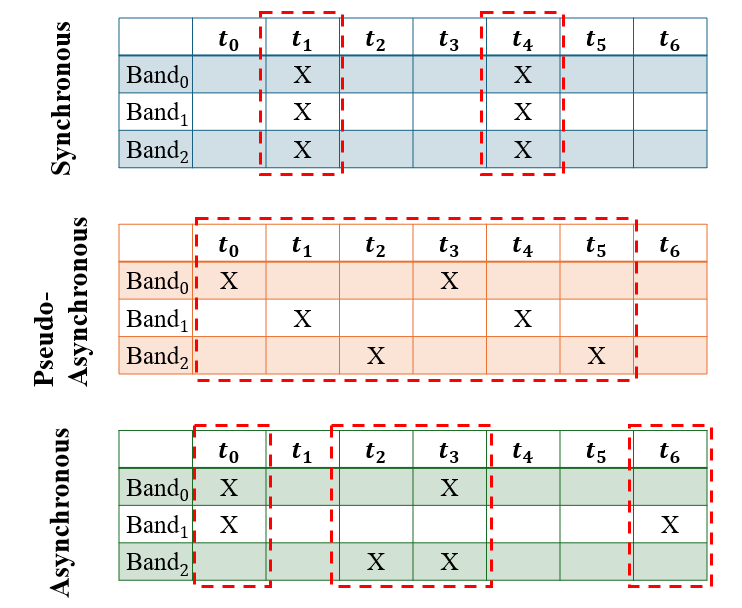}
    \caption{Comparison of the different sampling modes: synchronous, pseudo-asynchronous, and asynchronous. In this example, $N = 3$ and $L = 2$, so each of the three bands samples two observations from time $t_0, \dots, t_i$. The top table depicts the synchronous case, where each observation time sampled contains magnitude observations for all bands. The middle table depicts the pseudo-asynchronous case, where each band takes turns sampling at a periodic interval. The bottom table depicts the asynchronous case, where $L$ samples are randomly sampled for each band independently.}
    \label{fig:sync-vs-async}
\end{figure}

To handle a larger variety of survey data, our multiband framework is designed to ingest light curves in three distinct modes, each of which varies the temporal alignment of the observations across the $N$ bands. It should be noted that regardless of the mode used in our multiband model, the requirement that all relevant bands have valid observations and share the same time interval within the same light curve is strict. Inputs outside the shared time window are excluded; if a time step has valid observations in one band but missing values in another, the entire time step is excluded from the sequence. No imputation is performed. An example of the different modes is provided in Fig. \ref{fig:sync-vs-async}.

\paragraph{Synchronous mode} In the synchronous mode, the observed magnitudes from each band share the same observation time. This follows the format of surveys such as \texttt{MACHO} \citep{alcock_1993} and Gaia \citep{gaia_2016}.

That is, each magnitude $x_{i,j}^{(k)}$ of a specific band in the multiband input $Z^{(k)} = \left\{Z_0^{(k)}, \dots, Z_{N-1}^{(k)}\right\}$ shares the same time $t_{i,j}^{(k)}$ as exactly one other observed magnitude in each of the other $N$ bands. Here, $k$ indexes a light curve out of a total of $L$. The indices $i$ and $j$ represent the band and the observation within that band, respectively.

\paragraph{Pseudo-asynchronous mode} In the pseudo-asynchronous mode, the input sequence is constructed by interleaving observations from the $N$ bands in a fixed, periodic pattern. To test the performance of our model on survey data that are not necessarily synchronous, we introduce a pseudo-asynchronous sampling mode for synchronous datasets. In the case of \texttt{MACHO} where $N = 2$, the ingested sequence of magnitudes is defined by
\begin{equation}
    x_{i,j}^{(k)} = 
    \begin{cases}
    x_{0,j}^{(k)} & \text{if } j \text{ is even}, \\[1mm]
    x_{1,j}^{(k)} & \text{if } j \text{ is odd}.
    \end{cases}    
\end{equation}
More generally, for $N > 2$ the pattern cycles through the bands in a regular order.

\paragraph{Asynchronous mode} Given that the pseudo-asynchronous mode simulates irregular sampling of observations for the Alcock dataset, the asynchronous mode is our data sampling implementation for surveys that natively record observations in one band per time step, such as \texttt{ATLAS} \citep{tonry_2018}. Furthermore, the cadence of recording observations can be irregular as well.

In the asynchronous mode, the observations for each band are sampled independently without any coordinated time steps. For each light curve $k$ and band $i$, we randomly selected a subset of $L$ observations from the available $m_i$ data points,
\begin{equation*}
    {Z_{\text{sub}}}_i^{(k)} \subset Z_i^{(k)},    
\end{equation*}
where 
\begin{equation*}
    {Z_{\text{sub}}}_i^{(k)} = \left\{(t_{i,j'}^{(k)}, x_{i,j'}^{(k)}) : j' \in J \right\},
\end{equation*}
with $J$ being a uniformly random subset of $\{0,\ldots,m_i-1\}$ of cardinality $L$. In this setting, the observation times $t_{i,j'}^{(k)}$ are not aligned across bands. The encoder $\Phi_i$ processes each ${Z_{\text{sub}}}_i^{(k)}$ independently, and the resulting embeddings are then mixed via the mixing function $f$ to produce the unified representation. This mode requires that the model captures temporal dependences despite the absence of synchronized observations.

\section{Pre-processing of \texttt{MACHO} and \texttt{ATLAS} light curves}\label{sec:data}
Our work requires that the raw survey data undergo several pre-processing steps prior to model ingestion. This section describes the datasets used in this work, as well as the pre-processing steps required for our methodology.

\subsection{Surveys}

The \texttt{MACHO} \citep{alcock_1993} survey was conducted between 1992 and 1999 with the primary goal of detecting microlensing events toward the Magellanic Clouds and the Galactic bulge. In the process, the survey produced long-baseline optical light curves for millions of stars, observed primarily in two broad photometric bands. The resulting dataset has been widely used for studies of stellar variability and remains a benchmark resource for time-domain astronomy due to its dense temporal coverage and long observational baseline.

\texttt{ATLAS} \citep{tonry_2018} is an ongoing wide-field survey designed to detect near-Earth objects and optical transients. \texttt{ATLAS} observes the entire visible sky at high cadence using multiple photometric filters, producing large volumes of time series photometry. In addition to its primary mission, \texttt{ATLAS} data have proven valuable for variable star classification and other time-domain studies, particularly due to their multiband coverage and irregular sampling patterns.

\subsection{Data selection and pre-processing}
In our experiments, the multiband model uses unlabeled light curves for pre-training. The trained model is then fine-tuned on two different labeled datasets for downstream classification. Each dataset includes observation times and magnitudes across relevant photometric bands. As in the \texttt{Astromer}, we test the performance of our models by training over a balanced number of samples per class by random sampling (\( 20, 50, 100, \text{ and } 500 \)). The objective of these small sample sizes is to simulate a scenario where only a small labeled sample exists, making it impossible to perform pre-training. In this work, unlabeled light curves from the \texttt{MACHO} survey are used for large-scale pre-training, while labeled variable star datasets derived from \texttt{MACHO} (hereinafter \texttt{Alcock}) and \texttt{ATLAS} are used exclusively for fine-tuning and evaluation.

\subsection{Construction of an unlabeled dataset from the \texttt{MACHO} survey}

For the purposes of pre-training, we selected the same subset of fields (\texttt{`1'}, \texttt{`10'}, \texttt{`101'}, \texttt{`102'}, \texttt{`103'}, \texttt{`104'}) as the original \texttt{Astromer} implementation, aggregating to \num{1454792} light curves. This dataset is referred to as \texttt{MACHO} throughout this work. A proportion of $20\%$ of the light curves was reserved for the validation set, and a further $20\%$ for the test set to evaluate the model.

\begin{figure}
    \centering
    \includegraphics[scale=.51]{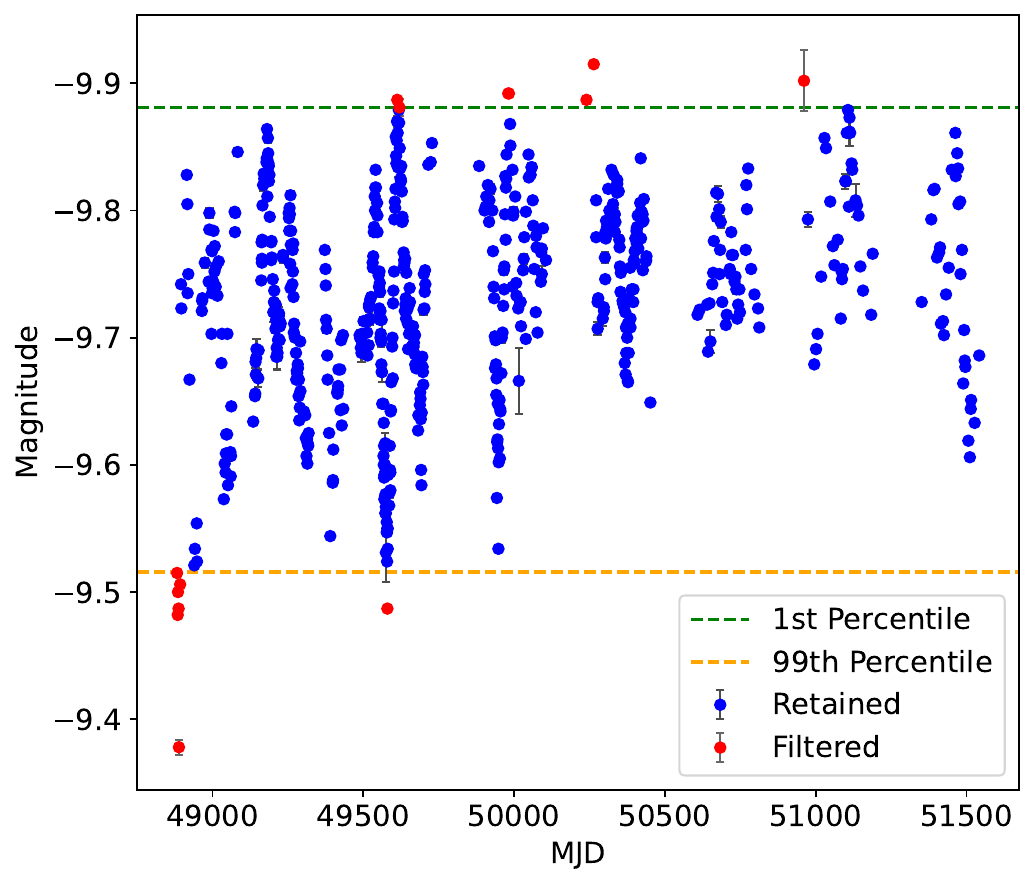}
    \caption{Example light curve showing observations in MJD and magnitude. The MJD is defined as $\text{MJD}=\text{Julian Date}-\num{2400000.5}$. Vertical error bars illustrate the photometric uncertainty, which are mostly smaller than the marker size. Brighter values appear lower on the vertical axis. Blue markers indicate observations retained after filtering, while red markers denote those excluded for falling outside the 1st and 99th percentile thresholds, indicated by dashed green and orange lines, respectively. Observations that were filtered out were not ingested by the neural networks.}
    \label{fig:per-lightcurve-filtering}
\end{figure}

\paragraph{Data cleaning and filtering}
We first cleaned the data by applying the same filtering protocol as the original \texttt{Astromer} paper. Specifically, light curves exhibiting white noise behavior were removed based on the following thresholds:
\begin{equation*}
    \text{Standard Deviation} > 0.1,\quad |\text{Skewness}| > 1,\quad|\text{Kurtosis}| > 10, 
\end{equation*}
which correspond to the second to fourth moment of the normal distribution.
Additional filtering measures included the removal of observations with negative uncertainties, which are present in the raw data.
To suppress the influence of extreme outliers, we removed observations falling outside the 1st and 99th percentiles of the magnitude distribution for each light curve. See Fig.~\ref {fig:per-lightcurve-filtering} for an illustration of the observation filtering process.

\paragraph{Signal-to-noise ratio (SNR) filtering}
After cleaning and filtering the data from statistical outliers, the dataset contained around \num{1400000} light curves. To shorten the pre-training time, we further reduce the number of light curves by extracting the \num{600000} highest SNR light curves to form the pre-training dataset. We compute approximate SNR using a simplified flux-based method adapted from \citep{carroll_2017}. For each light curve, the SNR was calculated using the interquartile range (IQR) of the flux values divided by the median flux error. Each observed magnitude $x$ was first transformed into its equivalent flux value $F$ by
\begin{equation}
    F = 10^{-0.4 x},
\end{equation}
and the corresponding flux uncertainty was calculated as
\begin{equation}
    \sigma_F = F \times 0.4 \times \ln (10) \times \sigma_x,
\end{equation}
where $\sigma_x$ is the magnitude uncertainty. The SNR value was then defined as the ratio of this IQR value to the median flux error $\tilde{\sigma_F}$:
\begin{equation}
    \operatorname{SNR} = \frac{\operatorname{IQR}}{\tilde{\sigma}_F}.
\end{equation}

\subsection{Labeled data}
As in the original \texttt{Astromer}, for downstream tasks, we incorporate two independently labeled catalogs: a subset of \texttt{MACHO} observations \citep[hereinafter \texttt{Alcock};][]{alcock_2003}, and \texttt{ATLAS} \citep{heinze_2018}. We chose these two different datasets to introduce some variation when testing our model's classification performance. Specifically, we investigate our model's performance under various conditions on the photometry and time observations.

\begin{figure*}[t]
    \centering
    \includegraphics[width=\textwidth]{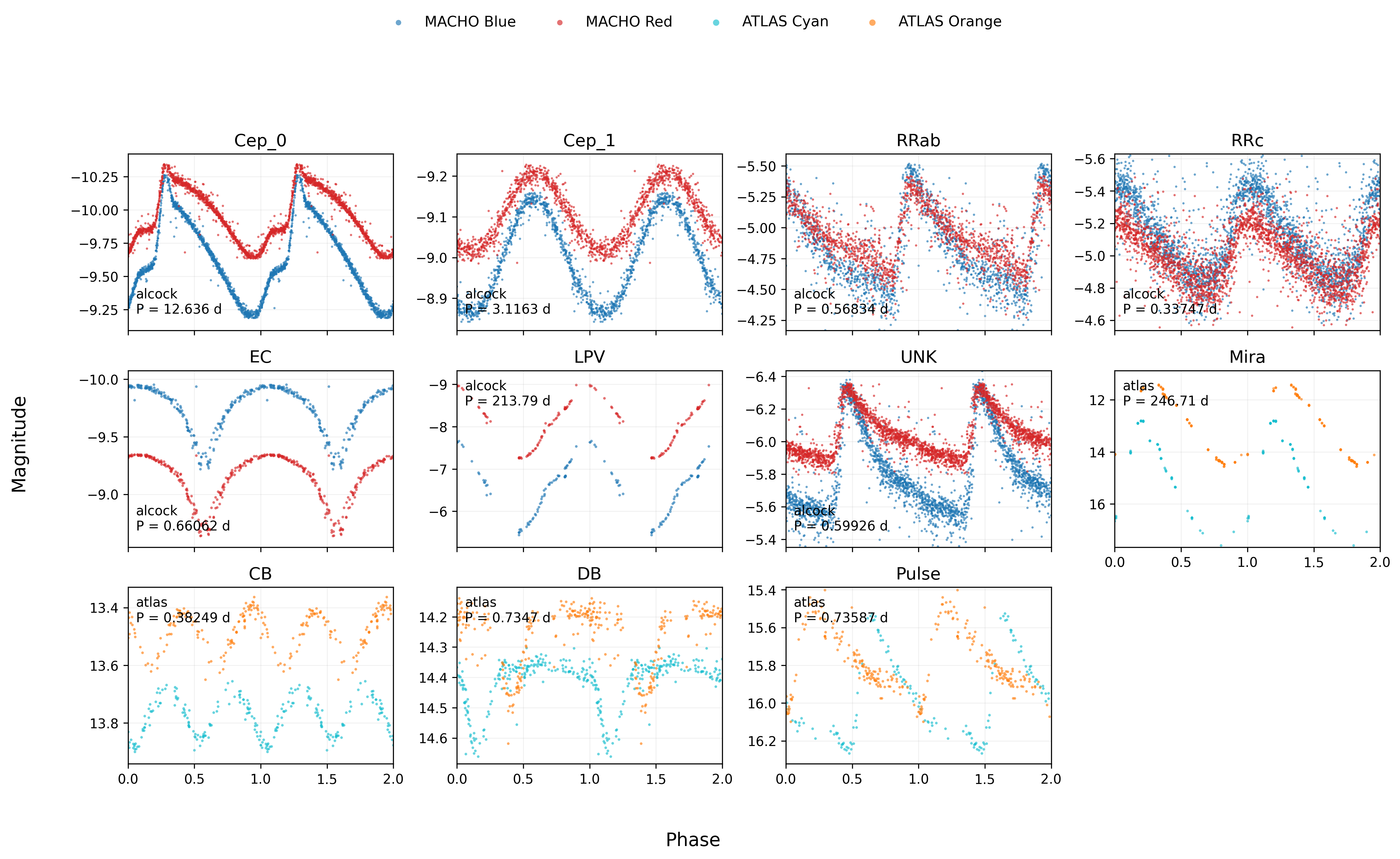}
    \caption{
    Representative phase-folded light curves for each labeled class used in the \texttt{Alcock} and \texttt{ATLAS} downstream classification datasets. Each panel shows one source, folded using the catalog period, with both photometric bands overlaid. \texttt{Alcock} panels show \texttt{MACHO} blue and red bands, while \texttt{ATLAS} panels show cyan and orange bands. The phase was repeated over two cycles for visual clarity, and the magnitude axis has been inverted so brighter observations appear higher.}
    \label{fig:lc-by-class}
\end{figure*}

The classes used in our experiments span the main families of periodic variability \citep[see][for a review]{catelan_2015}. RR Lyrae stars (RRab and RRc subtypes in the \texttt{Alcock} catalog) are horizontal-branch pulsators with periods below one day, distinguished by light-curve shape: RRab exhibit steep, sawtooth-like rises typical of fundamental-mode pulsation, while RRc display smoother, near-sinusoidal profiles from first-overtone pulsation. Classical Cepheids (Cep$_0$) and Type II Cepheids (Cep$_1$) are radial pulsators with periods of days to months, differing in stellar population and luminosity. Long-Period Variables (LPV in the \texttt{Alcock} catalog; Mira in the \texttt{ATLAS} catalog) are cool evolved giants with large-amplitude, quasi-periodic variations on timescales of hundreds of days. Eclipsing binaries (EC in \texttt{Alcock}; CB and DB in \texttt{ATLAS}) produce non-pulsational, geometric variability from mutual occultations, with light-curve morphologies (detached, semi-detached, contact) reflecting the binary configuration. The \texttt{ATLAS} Pulse class aggregates RR Lyrae, $\delta$-Scuti, and Cepheid pulsators into a single category. The UNK class in the \texttt{Alcock} catalog collects sources without a confident General Catalogue of Variable Stars \citep[GCVS;][]{samus_2017} classification. Representative phase-folded light curves for each labeled class are shown in Fig.~\ref{fig:lc-by-class}, with both photometric bands overlaid for each source.

\begin{figure}
    \centering
    \includegraphics[scale=.55]{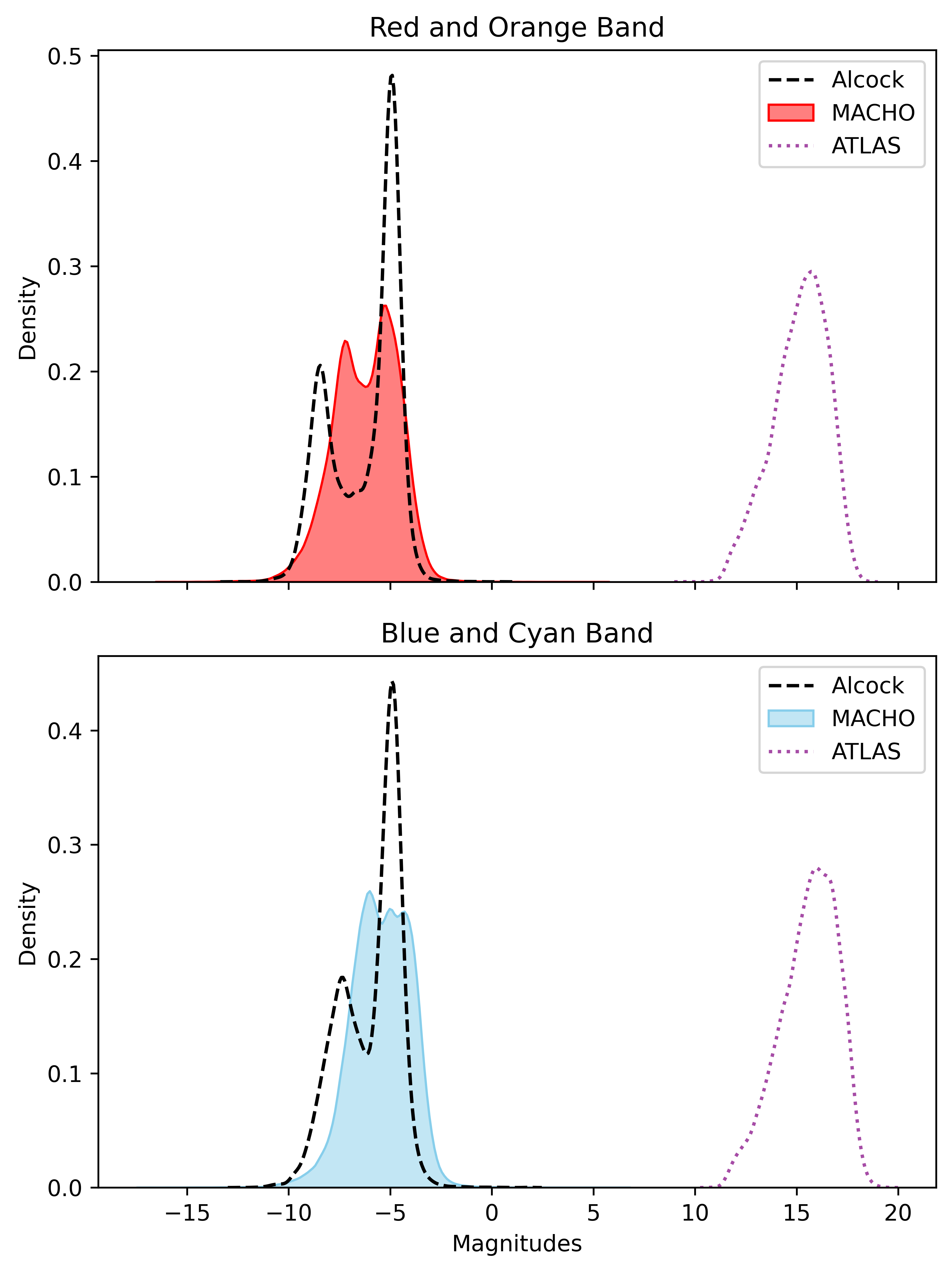}
    \caption{
    Magnitude distributions of the \texttt{MACHO}, \texttt{Alcock}, and \texttt{ATLAS} datasets across both bands used in our experiments. While \texttt{MACHO} and \texttt{Alcock} exhibit similar distributions due to their shared origin, the \texttt{ATLAS} distribution is offset due to differences in photometric zero points across surveys.
    "Blue" and "Red" refer to the MACHO optical bands, while "Cyan" and "Orange" refer to the ATLAS optical bands.
    Per-band mean and standard deviation of magnitudes: \texttt{MACHO} Red $\mu = -6.18$, $\sigma = 1.56$; \texttt{MACHO} Blue $\mu = -5.45$, $\sigma = 1.43$; \texttt{Alcock} Red $\mu = -6.24$, $\sigma = 1.68$; \texttt{Alcock} Blue $\mu = -5.97$, $\sigma = 1.46$; \texttt{ATLAS} Orange $\mu = 15.16$, $\sigma = 1.35$; \texttt{ATLAS} Cyan $\mu = 15.59$, $\sigma = 1.41$.
}
    \label{fig:macho-alcock-atlas-magn}
\end{figure}

\paragraph{\texttt{Alcock} dataset}
Our \texttt{Alcock} catalog comprises variable star light curves drawn from $30$ fields in the Large Magellanic Cloud (LMC), representing a subset of \texttt{MACHO} data. Following the original \texttt{Astromer}, the labels were updated to conform to modern variable-star classifications, following the nomenclature of the GCVS. The dataset includes \num{21444} light curves spanning six variable star classes (and one for unknown classes), as summarized in Tab.~\ref{tab:alcock}. It should be noted that we included the UNK class in our \texttt{Alcock} catalog, which was not used in the original \texttt{Astromer} paper. This category includes variable objects that did not fall in the other classes.

In Fig.~\ref{fig:macho-alcock-atlas-magn}, the \texttt{Alcock} and \texttt{MACHO} distributions exhibit strong overlap in the range of $-12$ to $0$ magnitudes. These negative magnitudes are a product of the original survey data, with no processing carried out in our pipeline to change this. The \texttt{Alcock} distribution exhibits multiple local maxima.

\begin{table}
\centering
\caption{Class distribution of variable stars in the \texttt{Alcock} catalog.}              
\label{tab:alcock}
\begin{tabular}{llr} 
\hline \hline
Tag    & Class Name         & \# of Sources \\ \hline
Cep\( _0 \) & Cepheid type I     & 1182          \\
Cep\( _1 \) & Cepheid type II    & 683           \\
EC     & Eclipsing binary   & 6824          \\
LPV    & Long period variable & 3046         \\
RRab   & RR Lyrae type ab   & 7397          \\
RRc    & RR Lyrae type c    & 1762          \\ 
UNK    & Unknown            & 550           \\ \hline
Total  &                    & \num{21444} \\ \hline
\end{tabular}
\end{table}

\paragraph{\texttt{ATLAS} dataset}
The variable star dataset consists of $4.7$ million candidate variable objects. We use the orange band to align our methodology with the original \texttt{Astromer} experiments, and the cyan band, which has a similar wavelength range to the blue \texttt{MACHO} band (see Tab.~\ref{tab:wavelength-range}). As most objects belong to the Dubious class, they are removed from our dataset. Furthermore, we apply the same data filtering and class grouping techniques, which result in a variable star catalog comprising \num{142024} light curves. The distribution of classes in \texttt{ATLAS} is provided in Tab.~\ref{tab:ATLAS}.

From Fig.~\ref{fig:macho-alcock-atlas-magn}, the \texttt{ATLAS} dataset forms a distinct peak centered around magnitudes 15 to 18, with no overlap with the \texttt{Alcock} and \texttt{MACHO} distributions. This offset is primarily due to differences in photometric zero points between the surveys, rather than intrinsic brightness differences of the observed sources. A detailed discussion of this calibration issue is beyond the scope of this section, but we note that such offsets are typical when comparing surveys with differing instrumentation and reduction pipelines. When fine-tuning our pre-trained models on the \texttt{ATLAS} dataset, we match the cyan band with the blue band from the \texttt{MACHO} dataset, and the orange band with \texttt{MACHO}’s red band, based on the wavelength similarity detailed in Table~\ref{tab:wavelength-range}.

\begin{table}
\centering
\caption{Class distribution of variable stars in the curated \texttt{ATLAS} dataset used for downstream classification.}              
\label{tab:ATLAS}
\begin{tabular}{llr} 
\hline \hline
Tag   & Class Name                                 & \# of Sources \\ \hline
CB    & Close binaries                             & \num{80368}         \\
DB    & Detached binary                            & \num{28895}         \\
Mira  & Mira variables                             & 7621          \\
Pulse & RR Lyrae, $\delta$-Scuti, Cepheids          & \num{25140}         \\ \hline
Total &                                            & \num{142024} \\ \hline
\end{tabular}
\end{table}

\begin{table}
\centering
\caption{Approximate wavelength coverage of photometric bands in the \texttt{MACHO} and \texttt{ATLAS} surveys.}
\label{tab:wavelength-range}
\begin{tabular}{llrr} 
\hline \hline
Survey  &  Filter Name  &  $\lambda_{min}$  &  $\lambda_{max}$  \\ 
  &    &  \r{A}  &  \r{A}  \\ 
\hline
 \texttt{MACHO}   &  Red          &  6300  &  7600  \\
 \texttt{MACHO}   &  Blue         &  4500  &  6300  \\
 \texttt{ATLAS}   &  Orange       &  5600  &  8200  \\
 \texttt{ATLAS}   &  Cyan         &  4200  &  6500  \\ \hline
\end{tabular}
\tablefoot{
This serves to justify our choice of band-matching in model fine-tuning.
}
\end{table}

The full lists of source identifiers, coordinates, and assigned class labels for the \texttt{Alcock} and \texttt{ATLAS} subsets used in this study are provided as plain-text files in the public code repository (see footnote in Sect.~\ref{sec:conclusion}). This enables direct reproduction of the train, validation, and test splits.

\subsection{Pre-processing}
Each survey has its own unique data schema. The difference in each survey's data schema requires the implementation of a data processing pipeline to standardize input data prior to model ingestion, as described below.

\paragraph{Fixed-length windowing}
Retaining the strategy in \texttt{Astromer} to limit the model to a manageable size, we adopt a fixed-length windowing approach, with $L = 200$. For light curves with more than $L$ observations, a random window of $200$ points is sampled, while sequences with fewer than 200 observations are zero-padded.

For the \texttt{Alcock} dataset, both the synchronous and pseudo-asynchronous sampling modes were obtained once the window length was determined. Note that the pseudo-asynchronous sampling mode requires a window length of $2L$ in order to obtain sufficient samples. On the other hand, the \texttt{ATLAS} dataset uses asynchronous sampling mode only.

\paragraph{Normalization and standardization}
Similar to the \texttt{Astromer} model's input pipeline, we normalize the values of the dataset after constructing the windows. Specifically, we apply per-light-curve mean subtraction (i.e., for each light curve we subtract its own mean magnitude, rather than min-max normalization or full standardization z-scoring, which would also rescale the amplitude). This preserves the morphology of each light curve while removing absolute zero-point differences between sources. Absolute magnitudes are not ingested as separate supplementary features; the model learns from the temporal and amplitude structure of each light curve relative to its own mean, consistent with the original \texttt{Astromer} pipeline. Following the data pre-processing step, each dataset was then partitioned in preparation for model training, with a 60\% training, 20\% validation, and 20\% test split.

\section{Results}\label{sec:results}

Our experiments were carried out using the Cannon Cluster\footnote{\url{https://www.rc.fas.harvard.edu/}}, a high-performance computing system at the Faculty of Arts and Sciences at Harvard University. All experiments were executed on NVIDIA A100 GPUs, each with 80GB of VRAM. The CPU used was the $3$rd Generation Intel Xeon Scalable Ice Lake-SP with $64$ cores per node, with each node having $1$TB of memory. We implemented \texttt{Multiband Astromer} using Tensorflow 2 \citep{tensorflow2015-whitepaper}.

We report the quantitative performance of the proposed architectures, with all metrics averaged over five independent runs. Throughout this section, we use $\mathcal{E}$ to denote the RMSE. Unless otherwise stated, we report deviations as standard deviations $\sigma$. 
To provide a baseline to assess the performance of our multiband models, we ran the original \texttt{Astromer} models under the same hardware and data conditions as our multiband models for comparison. For convenience, we describe our single-band models with a similar short-hand notation as the multiband models by the hyphen-separated architecture prefix of "\texttt{Single}" and suffix of band color (i.e., \texttt{Blue} or \texttt{Red}).

\subsection{Pre-training}

\begin{figure}[hptb]
    \centering
    \includegraphics[scale=.29]{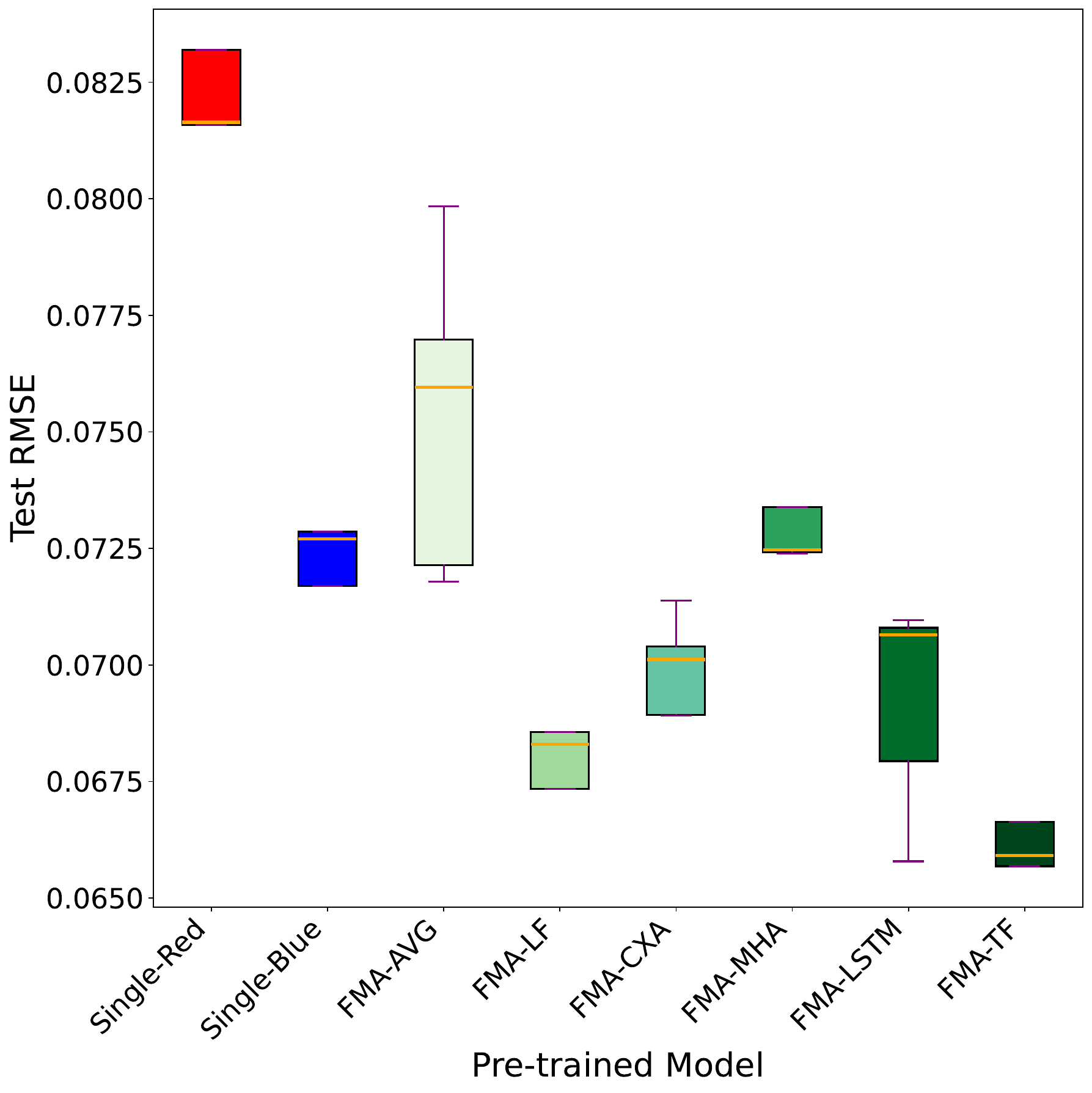}
    \caption{Median of the RMSE over five runs for each model pre-trained on our unlabeled \texttt{MACHO} dataset using masked reconstruction. Boxes show IQR; whiskers indicate data points within $1.5 \times$ IQR. The \texttt{Single-Red} model performs the worst overall. The \texttt{FMA-TF} model performs the best.}
    \label{fig:pt-boxplot}
\end{figure}

We omitted \texttt{SMA} from pre-training results since it reuses independently trained single-band encoders and does not undergo multiband pre-training. The model training was configured using an early stopping patience of 40 epochs. Figure~\ref{fig:pt-boxplot} summarizes the RMSE performance across architectures over five runs.

The \texttt{Single-Blue} model achieves a lower error ($\mathcal{E}_{\texttt{Single-Blue}} = 0.073$) than the \texttt{Single-Red} model ($\mathcal{E}_{\texttt{Single-Red}} = 0.082$). Among \texttt{FMA} variants, the TF mixing strategy achieves the best performance ($\mathcal{E}_{\texttt{FMA-TF}} = 0.066$), while AVG, the simplest approach, performs worse ($\mathcal{E}_{\texttt{FMA-AVG}} = 0.076$), highlighting the benefit of fusion methods with higher parameter counts. In terms of stability, \texttt{FMA-TF} and \texttt{FMA-CXA} mixing show the lowest variance across runs, while \texttt{FMA-AVG} has the highest, as evident in the whisker spread of Fig.~\ref{fig:pt-boxplot}.

Training time varies with model complexity. Single-band models typically converge in 20–65 hours. In contrast, \texttt{FMA} models with complex fusion layers (e.g., TF, CXA, LSTM) often exceed the cluster’s 72-hour limit. Simpler models such as \texttt{FMA-AVG} converge faster thanks to the smaller number of hyperparameters in these models. Table~\ref{tab:pt-times} summarizes the training durations, illustrating the trade-off between accuracy and computational cost.

\begin{table}
\centering
\caption{Pre-training running times recorded across five experimental runs (in hours).}
\label{tab:pt-times}
\begin{tabular}{p{2cm}llllllll}
\hline \hline
        Model & Run 1 & Run 2 & Run 3 & Run 4 & Run 5 \\ \hline
        \texttt{Single-Blue} & >72 & 64.6 & 23.9 & 67.3 & 37.8 \\
        \texttt{Single-Red} & 38.1 & 28.9 & 29.5 & 46.8 & 40.4 \\
        \raggedright \texttt{FMA-AVG} & >72 & >72 & 55.0 & 59.0 & 61.9 \\
        \texttt{FMA-LF} & 52.8 & >72 & >72 & 55.9 & >72 \\
        \raggedright \texttt{FMA-CXA} & >72 & >72 & >72 & >72 & >72 \\
        \raggedright \texttt{FMA-MHA} & >72 & 67.5 & 58.3 & >72 & >72 \\
        \raggedright \texttt{FMA-LSTM} & >72 & 62.5 & >72 & >72 & >72 \\
        \raggedright \texttt{FMA-TF} & >72 & >72 & >72 & >72 & >72 \\ \hline
    \end{tabular}
    \tablefoot{
Runs are automatically stopped at 72 hours.
}
\end{table}

\subsection{Fine-tuning performance}
Fine-tuning in our setup refers to a second stage of unsupervised masked reconstruction, applied to a smaller, labeled catalog. Although the datasets contain class labels, these are not used during fine-tuning. The loss function remains the same as in pre-training (Eq.~\ref{eq:rmse-loss}), and all model components, encoders, mixing layers, and decoder, are trained end-to-end.

Sampling by class is used solely to balance the number of light curves per class; the labels themselves are never included in the loss calculation. This design mirrors the methodology of the original \texttt{Astromer} study. Training early stopping was configured with a patience of 30 epochs on validation loss.

We fine-tuned each model on the \texttt{Alcock} and \texttt{ATLAS} datasets using subsets with 20, 50, 100, and 500 samples per class as well as the full datasets. In addition, we evaluated the performance under three observational sampling modes: synchronous, pseudo-asynchronous, and asynchronous under the fixed 100-samples-per-class test set.

Table~\ref{tab:ft-rmse-merged} shows the fine-tuning RMSE on the \texttt{Alcock} dataset under synchronous sampling. The top panel reports results for single-band models: the \texttt{Single-Red} and \texttt{Single-Blue} models have minimal performance differences with each other and across the different datasets in the $\mathcal{E}_R, \mathcal{E}_B \in [0.10, 0.12]$ interval. However, single-band models remain less competitive compared to the multiband architectures, particularly when the dataset sizes increase (e.g., $\mathcal{E}_{\texttt{FMA-TF}}=0.07$ for the entire dataset).

The bottom panel in Tab.~\ref{tab:ft-rmse-merged} compares multiband models. As expected, performance generally improves with more training data. Models using \texttt{FMA} with mixing strategies of TF, LSTM, and LF achieve the lowest errors ($\sim$0.07 at 500+ samples), while \texttt{SMA} variants, particularly those using AVG and MHA, exhibit higher RMSE ($\geq 0.18$) at lower dataset sizes, while the performance differences between \texttt{SMA} and \texttt{FMA} are much narrower at higher dataset sizes (between 0.06 and 0.10 across all models for the entire dataset).

\begin{table}[ht]
\centering
\caption{Average test RMSE for fine-tuning experiments across datasets and training settings.}

\begin{tabular}{lccccc}
\hline \hline
Model & \multicolumn{5}{c}{Samples Per Class} \\
& 20 & 50 & 100 & 500 & All \\
\hline

\multicolumn{6}{l}{\texttt{Alcock} — Synchronous} \\
\hline

\quad \texttt{Single-Red}   & 0.12 & 0.12 & 0.11 & 0.11 & 0.11 \\
\quad \texttt{Single-Blue}  & 0.12 & 0.12 & 0.12 & 0.12 & 0.10 \\

\hline
\quad \texttt{SMA-AVG}   & 0.20 & 0.26 & 0.21 & 0.18 & 0.08 \\
\quad \texttt{SMA-LF}    & 0.14 & 0.20 & 0.12 & 0.09 & 0.07 \\
\quad \texttt{SMA-CXA}   & 0.26 & 0.23 & 0.23 & 0.21 & 0.08 \\
\quad \texttt{SMA-MHA}   & 0.26 & 0.26 & 0.22 & 0.22 & 0.07 \\
\quad \texttt{SMA-LSTM}  & 0.18 & 0.23 & 0.13 & 0.15 & 0.09 \\
\quad \texttt{SMA-TF}    & 0.20 & 0.21 & 0.19 & 0.10 & 0.08 \\

\hline
\quad \texttt{FMA-AVG}   & 0.12 & 0.12 & 0.11 & 0.11 & 0.10 \\
\quad \texttt{FMA-LF}    & 0.08 & 0.08 & 0.08 & 0.07 & 0.07 \\
\quad \texttt{FMA-CXA}   & 0.11 & 0.11 & 0.10 & 0.10 & 0.08 \\
\quad \texttt{FMA-MHA}   & 0.11 & 0.11 & 0.11 & 0.11 & 0.10 \\
\quad \texttt{FMA-LSTM}  & 0.08 & 0.08 & 0.08 & 0.07 & 0.06 \\
\quad \texttt{FMA-TF}    & 0.08 & 0.08 & 0.08 & 0.07 & 0.07 \\

\hline
\multicolumn{6}{l}{\texttt{Alcock} — Pseudo-asynchronous} \\
\hline

\quad \texttt{SMA-AVG}  & 0.23 & 0.26 & 0.23 & 0.17 & 0.11 \\
\quad \texttt{SMA-LF}   & 0.14 & 0.17 & 0.14 & 0.12 & 0.09 \\
\quad \texttt{SMA-CXA}  & 0.25 & 0.26 & 0.26 & 0.24 & 0.14 \\
\quad \texttt{SMA-MHA}  & 0.27 & 0.26 & 0.19 & 0.20 & 0.13 \\
\quad \texttt{SMA-LSTM} & 0.24 & 0.23 & 0.20 & 0.13 & 0.10 \\
\quad \texttt{SMA-TF}   & 0.17 & 0.26 & 0.22 & 0.13 & 0.10 \\

\hline
\quad \texttt{FMA-AVG}  & 0.12 & 0.12 & 0.12 & 0.12 & 0.11 \\
\quad \texttt{FMA-LF}   & 0.10 & 0.10 & 0.10 & 0.09 & 0.09 \\
\quad \texttt{FMA-CXA}  & 0.12 & 0.11 & 0.11 & 0.11 & 0.10 \\
\quad \texttt{FMA-MHA}  & 0.12 & 0.12 & 0.12 & 0.12 & 0.11 \\
\quad \texttt{FMA-LSTM} & 0.08 & 0.08 & 0.08 & 0.08 & 0.08 \\
\quad \texttt{FMA-TF}   & 0.08 & 0.08 & 0.08 & 0.08 & 0.07 \\

\hline
\multicolumn{6}{l}{\texttt{ATLAS} — Asynchronous} \\
\hline

\quad \texttt{SMA-AVG}  & 0.40 & 0.42 & 0.36 & 0.34 & 0.08 \\
\quad \texttt{SMA-LF}   & 0.28 & 0.22 & 0.20 & 0.11 & 0.08 \\
\quad \texttt{SMA-CXA}  & 0.39 & 0.38 & 0.37 & 0.38 & 0.08 \\
\quad \texttt{SMA-MHA}  & 0.38 & 0.36 & 0.37 & 0.35 & 0.23 \\
\quad \texttt{SMA-LSTM} & 0.40 & 0.38 & 0.31 & 0.28 & 0.07 \\
\quad \texttt{SMA-TF}   & 0.42 & 0.33 & 0.25 & 0.24 & 0.08 \\

\hline
\quad \texttt{FMA-AVG}  & 0.25 & 0.25 & 0.23 & 0.21 & 0.08 \\
\quad \texttt{FMA-LF}   & 0.19 & 0.17 & 0.18 & 0.14 & 0.08 \\
\quad \texttt{FMA-CXA}  & 0.25 & 0.23 & 0.22 & 0.22 & 0.09 \\
\quad \texttt{FMA-MHA}  & 0.23 & 0.23 & 0.22 & 0.19 & 0.08 \\
\quad \texttt{FMA-LSTM} & 0.20 & 0.19 & 0.19 & 0.14 & 0.08 \\
\quad \texttt{FMA-TF}   & 0.18 & 0.18 & 0.16 & 0.14 & 0.07 \\

\hline
\end{tabular}
\tablefoot{The lowest value in each column within a dataset block corresponds to the best-performing model for that setting.}
\label{tab:ft-rmse-merged}
\end{table}

When using pseudo-asynchronous sampling (Tab.~\ref{tab:ft-rmse-merged}), overall results follow the synchronous case: \texttt{FMA} remains more robust and accurate across all dataset sizes (with an RMSE between 0.07 and 0.12), while \texttt{SMA} exhibits larger variance and higher RMSE, especially at small sample sizes (e.g., $\mathcal{E}_{\texttt{SMA-AVG}}=0.23$ for 20 samples per class). The performance difference narrows slightly as training size increases, but expressive \texttt{FMA} variants (e.g., TF, LSTM) still dominate, achieving $\mathcal{E} \leq 0.08$ on the full dataset as training size increases and \texttt{SMA} models improve in performance, although \texttt{FMA} models still perform marginally better, with the RMSE ranging between 0.07 and 0.11, while the \texttt{SMA} models range between 0.09 and 0.14.

Results on the \texttt{ATLAS} dataset, which uses asynchronous sampling, are shown in Tab.~\ref{tab:ft-rmse-merged} and follow the results of the pseudo-asynchronous case: \texttt{FMA} models outperform \texttt{SMA}, especially at lower sample sizes, where \texttt{SMA} variants have RMSE values ranging between 0.28 and 0.42, while \texttt{FMA} models remain under 0.25 for 20 samples per class. When more data are available, all models improve and differences narrow mostly into the range of 0.07 to 0.09. However, the performance on \texttt{ATLAS} is generally worse than \texttt{Alcock}, likely due to distributional shifts between both pre-training datasets.

In summary, fine-tuning confirms that \texttt{FMA} models, particularly those with TF and LSTM mixing, yield lower errors and greater consistency at smaller datasets, while the difference with \texttt{SMA} models becomes negligible when more data are available. \texttt{SMA} models benefit from multiband input when more data are available, but remain limited in scarce data environments due to their lack of joint pre-training.

\subsection{Classification performance}

We evaluate the performance of our multiband model in the supervised setting by replacing the masked reconstruction decoder with a classification head from the original \texttt{Astromer}. All encoders, fusion layers, and classification parameters are trained end-to-end using cross-entropy loss (Eq. \ref{eq:cross-entropy}). Performance is reported as the macro F1-score (Eq. \ref{eq:f1-score}) across multiple labeled datasets and sampling modes.

We test models on both the \texttt{Alcock} and \texttt{ATLAS} datasets using training subsets of 20, 50, 100, and 500 samples per class, as well as the full dataset. Early stopping was configured with a patience of 30 epochs on validation loss. We report results under synchronous, pseudo-asynchronous, and asynchronous sampling using the fixed 100-samples-per-class test set.

\paragraph{F1-score across sampling modes}
Table~\ref{tab:ft-f1-merged} shows F1-scores for the \texttt{Alcock} dataset under synchronous sampling. As expected, all models improve with more training data, with F1-scores ranging from 0.15 to 0.48 for 20 samples per class and from 0.62 to 0.74 for the full dataset. \texttt{FMA} models outperform their \texttt{SMA} counterparts across most sample sizes and mixing strategies, achieving scores up to $\text{F1-score} \sim 0.7$ at fewer samples per class datasets (i.e., $F_{1,\texttt{FMA}} \in [0.38, 0.48]$ versus $F_{1,\texttt{SMA}} \in [0.15,0.40]$ for 20 samples per class), but the difference becomes negligible with larger dataset sizes. LSTM, TF, and LF perform best in both \texttt{SMA} and \texttt{FMA} variants, as evidenced by being the top performers across all dataset sizes.

\begin{table}[ht]
\centering
\caption{Average classification F1-scores for fine-tuning experiments across datasets and training settings.}

\begin{tabular}{lccccc}
\hline \hline
Model & \multicolumn{5}{c}{Samples Per Class} \\
& 20 & 50 & 100 & 500 & All \\
\hline

\multicolumn{6}{l}{\texttt{Alcock} — Synchronous} \\
\hline

\quad \texttt{Single-Red}                  & 0.34 & 0.47 & 0.50 & 0.63 & 0.64 \\
\quad \texttt{Single-Blue}                 & 0.35 & 0.41 & 0.50 & 0.59 & 0.62 \\

\hline
\quad \texttt{SMA-AVG}                     & 0.39 & 0.28 & 0.57 & 0.64 & 0.71 \\
\quad \texttt{SMA-LF}                      & 0.39 & 0.50 & 0.66 & 0.69 & 0.71 \\
\quad \texttt{SMA-CXA}                     & 0.27 & 0.50 & 0.60 & 0.71 & 0.71 \\
\quad \texttt{SMA-MHA}                     & 0.15 & 0.24 & 0.34 & 0.40 & 0.70 \\
\quad \texttt{SMA-LSTM}                    & 0.40 & 0.54 & 0.69 & 0.71 & 0.74 \\
\quad \texttt{SMA-TF}                      & 0.36 & 0.50 & 0.63 & 0.72 & 0.73 \\

\hline
\quad \texttt{FMA-AVG}                     & 0.39 & 0.52 & 0.61 & 0.67 & 0.72 \\
\quad \texttt{FMA-LF}                      & 0.48 & 0.57 & 0.66 & 0.72 & 0.72 \\
\quad \texttt{FMA-CXA}                     & 0.42 & 0.47 & 0.60 & 0.70 & 0.70 \\
\quad \texttt{FMA-MHA}                     & 0.38 & 0.51 & 0.55 & 0.66 & 0.71 \\
\quad \texttt{FMA-LSTM}                    & 0.42 & 0.51 & 0.60 & 0.73 & 0.72 \\
\quad \texttt{FMA-TF}                      & 0.47 & 0.57 & 0.66 & 0.71 & 0.71 \\

\hline

\multicolumn{6}{l}{\texttt{Alcock} — Pseudo-asynchronous} \\
\hline

\quad \texttt{SMA-AVG}                     & 0.43 & 0.36 & 0.50 & 0.63 & 0.66 \\
\quad \texttt{SMA-LF}                      & 0.40 & 0.56 & 0.66 & 0.73 & 0.71 \\
\quad \texttt{SMA-CXA}                     & 0.27 & 0.26 & 0.34 & 0.45 & 0.69 \\
\quad \texttt{SMA-MHA}                     & 0.26 & 0.33 & 0.58 & 0.56 & 0.56 \\
\quad \texttt{SMA-LSTM}                    & 0.24 & 0.28 & 0.50 & 0.69 & 0.69 \\
\quad \texttt{SMA-TF}                      & 0.43 & 0.30 & 0.52 & 0.69 & 0.70 \\

\hline
\quad \texttt{FMA-AVG}                     & 0.38 & 0.51 & 0.57 & 0.63 & 0.69 \\
\quad \texttt{FMA-LF}                      & 0.47 & 0.58 & 0.61 & 0.70 & 0.70 \\
\quad \texttt{FMA-CXA}                     & 0.46 & 0.54 & 0.62 & 0.71 & 0.72 \\
\quad \texttt{FMA-MHA}                     & 0.43 & 0.47 & 0.57 & 0.69 & 0.69 \\
\quad \texttt{FMA-LSTM}                    & 0.47 & 0.59 & 0.63 & 0.72 & 0.72 \\
\quad \texttt{FMA-TF}                      & 0.49 & 0.62 & 0.63 & 0.70 & 0.71 \\

\hline

\multicolumn{6}{l}{\texttt{ATLAS} — Asynchronous} \\
\hline

\quad \texttt{SMA-AVG}                     & 0.32 & 0.55 & 0.67 & 0.78 & 0.93 \\
\quad \texttt{SMA-LF}                      & 0.64 & 0.72 & 0.80 & 0.85 & 0.92 \\
\quad \texttt{SMA-CXA}                     & 0.38 & 0.47 & 0.59 & 0.78 & 0.92 \\
\quad \texttt{SMA-MHA}                     & 0.63 & 0.56 & 0.65 & 0.81 & 0.91 \\
\quad \texttt{SMA-LSTM}                    & 0.70 & 0.72 & 0.83 & 0.80 & 0.93 \\
\quad \texttt{SMA-TF}                      & 0.26 & 0.59 & 0.79 & 0.85 & 0.92 \\

\hline
\quad \texttt{FMA-AVG}                     & 0.65 & 0.65 & 0.76 & 0.86 & 0.92 \\
\quad \texttt{FMA-LF}                      & 0.78 & 0.65 & 0.77 & 0.86 & 0.92 \\
\quad \texttt{FMA-CXA}                     & 0.76 & 0.70 & 0.77 & 0.83 & 0.92 \\
\quad \texttt{FMA-MHA}                     & 0.77 & 0.65 & 0.81 & 0.85 & 0.92 \\
\quad \texttt{FMA-LSTM}                    & 0.67 & 0.74 & 0.81 & 0.86 & 0.93 \\
\quad \texttt{FMA-TF}                      & 0.81 & 0.74 & 0.75 & 0.85 & 0.92 \\

\hline
\end{tabular}
\tablefoot{The highest value in each column within a dataset block corresponds to the best-performing model for that setting.}
\label{tab:ft-f1-merged}
\end{table}

Under pseudo-asynchronous sampling shown in Tab.~\ref{tab:ft-f1-merged}, results mirror the synchronous case. \texttt{FMA} models remain more stable across runs and outperform the \texttt{SMA} variants at fewer samples per class datasets (i.e., $F_{1,\texttt{FMA}} \in [0.38,0.49]$ versus $F_{1,\texttt{SMA}} \in [0.24,0.43]$ for 20 samples per class). Some \texttt{SMA} strategies (e.g., TF and LSTM) perform comparably to \texttt{FMA} at large sample sizes, but show higher deviations across experiments at 20 and 50 samples per class (e.g., $F_{1,\texttt{SMA-LSTM}}=0.24$, while $F_{1,\texttt{FMA-LSTM}}=0.47$ for 20 samples per class dataset).

Table~\ref{tab:ft-f1-merged} shows results on the \texttt{ATLAS} dataset under asynchronous sampling. As with \texttt{Alcock}, \texttt{FMA} models dominate, especially at smaller data regimes (e.g., $F_{1,\texttt{FMA}} \in [0.65,0.81]$ versus $F_{1,\texttt{SMA}} \in [0.26,0.70]$). \texttt{SMA} variants show large variability and struggle to generalize with limited data. When more is available, performance across architectures converges, with all models reporting results between 0.91 and 0.93 when trained on the entire dataset.

\subsection{Detailed class analysis}

To better understand model behavior, we analyze the \texttt{FMA-TF} model over five runs at 500 samples per class. It ranks among the top-performing configurations across datasets and training regimes. Although there is not one single best model by F1-score, this strategy remains within a few percentage points of the highest-performing configurations in nearly all settings. For the \texttt{Alcock} synchronous dataset, its F1-score is within $0.01$–$0.06$ of the top strategy across $20$–$500$ samples per class and all-sample conditions, being the best performer at $50$ samples per class. Under pseudo-asynchronous training, it achieves the highest score at both $20$ and $50$ samples per class and remains within $0.03$ of the best in all other cases. For the \texttt{ATLAS} asynchronous dataset, it attains the top performance at $20$ and $50$ samples per class, and lies within $0.01$–$0.08$ of the best elsewhere.

\begin{figure}[h!]
\centering
\includegraphics[scale=.44]{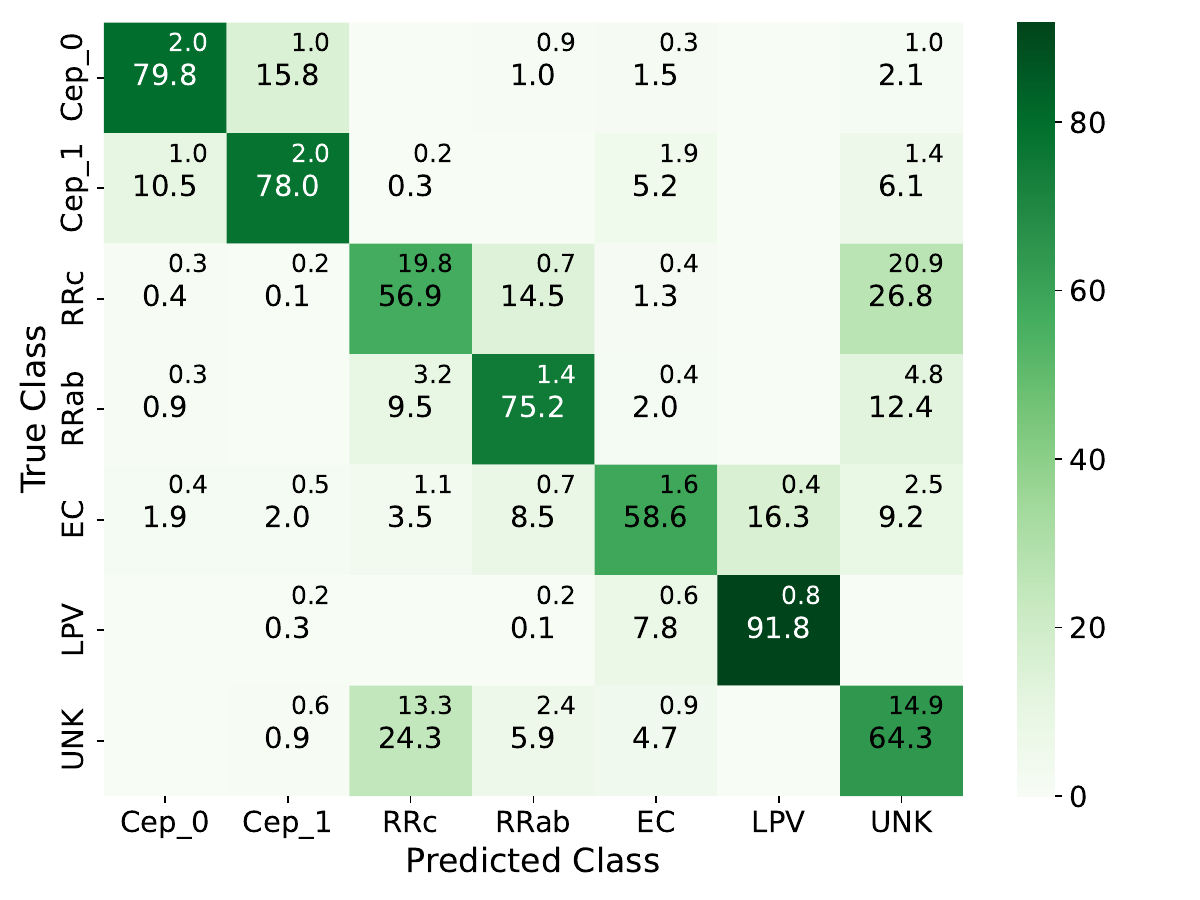}
\caption{Confusion matrix for \texttt{FMA-TF} on the \texttt{Alcock} dataset, pseudo-asynchronous mode. The accuracy is reported as the larger number centered within a square, while the smaller number at the top-right corner of each square reports the standard deviation.}
\label{fig:cm-async-alcock-TF-single}
\end{figure}

\paragraph{\texttt{Alcock} dataset (pseudo-asynchronous)}
Figure~\ref{fig:cm-async-alcock-TF-single} shows the confusion matrix for the \texttt{Alcock} dataset. The LPV class is most accurately predicted (91.8\%), with low variance. RRc and EC are the most challenging classes, with the lowest model successful prediction accuracies near 57–59\%.

Bidirectional confusion between Cep$_0$ and Cep$_1$ is common, as is leakage from UNK to RRc. These patterns reflect astrophysical similarities; in particular, RRc stars exhibit overtone pulsations with lower amplitudes, which produce less distinctive light curve profiles and can lead to classification ambiguity.

\begin{figure}[h!]
\centering
\includegraphics[scale=.44]{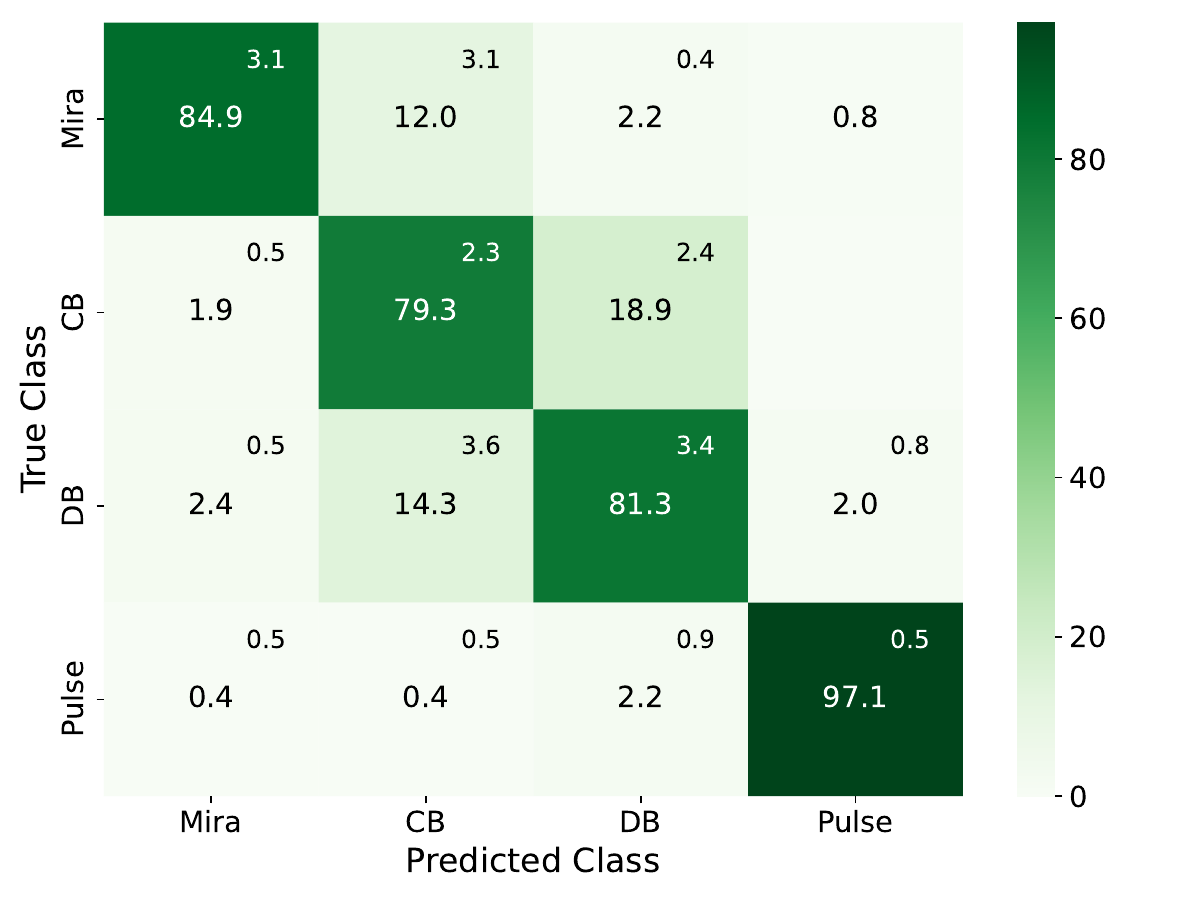}
\caption{Confusion matrix for \texttt{FMA-TF} on the \texttt{ATLAS} dataset, asynchronous mode. The accuracy is reported as the larger number centered within a square, while the smaller number at the top-right corner of each square reports the standard deviation. Pulse class is most accurately predicted.}
\label{fig:cm-async-atlas-TF-single}
\end{figure}

\paragraph{\texttt{ATLAS} dataset (asynchronous)}
Figure~\ref{fig:cm-async-atlas-TF-single} presents the confusion matrix for the \texttt{ATLAS} dataset. The Pulse class achieves 97.1\% accuracy with minimal variance (0.5\%). In contrast, CB and DB have lower accuracy (79–81\%) and frequent mutual misclassifications. The CB light curves are often mistaken for Miras (12\%), though the asymmetric confusion is rare (2\%), which is likely tied to smoother CB profiles compared to sharp periodic features in the Pulse class.

\section{Conclusion}\label{sec:conclusion}

We have introduced \texttt{Multiband Astromer}, an extension of the original single-band \texttt{Astromer}. The extension is designed to incorporate information across bands for multiband light curve analysis. By characterizing correlations across asynchronously sampled photometric bands, our approach improves on both reconstruction (up to nearly $50\%$ improvement) and classification (approximately ten percentage points) performance over single-band baselines across our constructed labeled and unlabeled datasets from the \texttt{MACHO} and \texttt{ATLAS} surveys.

Our experiments show that the multiband architectures \texttt{FMA} and \texttt{SMA} improve classification with an increase in F1-score of approximately ten percentage points compared to single-band models under synchronous sampling in our labeled \texttt{Alcock} dataset (i.e., the F1-score for the single-band models were between $0.62$ and $0.64$, while the F1-score for the \texttt{SMA} models were between $0.70$ and $0.74$, and the F1-score for the \texttt{FMA} models were between $0.70$ and $0.72$ when trained on the entire dataset). We evaluated a variety of embedding fusion strategies, including simple averaging, learnable fusion via multilayer perceptron, and layers with higher parameter counts such as cross-attention and transformer blocks. Of these, the more complex methods (e.g., LSTM or TF mixing) yielded the best classification performance across all experiments (as shown by the highest values in Table \ref{tab:ft-f1-merged}), highlighting the value of explicitly characterizing cross-band dependences.

We also compared the two model architectures: \texttt{SMA} and \texttt{FMA}. The \texttt{SMA} framework enables reuse of independently pre-trained single-band encoders with negligible integration cost. However, its decoupled training limits coordination across bands. In contrast, \texttt{FMA} jointly pre-trains all bands under a shared loss, reducing the reconstruction error from 0.10 to 0.07 (TF mixing for pseudo-asynchronous sampling with the \texttt{Alcock} dataset) but often doubling the pre-training time. This exposes a key trade-off between the flexibility of reusing different single-band encoders and the representational power of joint multiband training.

Even when data are scarce, multiband models consistently outperform their single-band counterparts (i.e., for 20 samples per class, the F1-score for the single-band models achieved up to $0.35$, while the multiband models had F1-scores of up to $0.48$ for synchronous sampling). Model layers with higher parameter counts generally yield better accuracy and exhibit greater stability across sampling regimes. Notably, there are minimal differences between the synchronous and asynchronous sampling modes, suggesting that strict temporal alignment is not essential to leverage multiband information.

These results suggest that \texttt{SMA}-style models may be well-suited to scalable pipelines in large filter-rich surveys, whereas \texttt{FMA} may offer advantages when data are scarce, imbalanced, or noisy. Ultimately, flexible multiband architectures present a promising direction for general-purpose time series models in astronomy.

Future work may expand to additional datasets or explore architectural refinements, such as learned positional encodings or spectral attention, which conditions attention weights on band identity in order to improve cross-band alignment. As wide-field surveys expand the volume and variety of multiband observations, the ability to integrate asynchronous multivariate time series will become increasingly central to efficiently processing and accelerating discoveries from massive astrophysical datasets.

Large time-domain surveys, ranging from early foundational projects such as \texttt{MACHO} to contemporary and next-generation facilities such as Gaia and LSST, operate at a massive scale, with millions of multiband light curves, irregular cadences, asynchronous observations, and limited labeled data. The \texttt{Multiband Astromer} framework is explicitly designed for the analysis of periodic objects. Further research is required to evaluate its application in stochastic or transient objects. The \texttt{SMA} model enables modular, reusable pre-training of single-band encoders, facilitating distributed training, model sharing across surveys, and reduced computational and environmental cost. The \texttt{FMA} model, while more computationally demanding, allows coordinated learning of the cross-band correlations when sufficient multiband coverage is available. The range of embedding mixing strategies explored reflects practical trade-offs between scalability and representational power, while the explicit support for asynchronous sampling mirrors the dominant data acquisition mode of contemporary wide-field surveys. Together, these design choices make \texttt{Multiband Astromer} well suited for deployment in large-scale time-domain survey pipelines.

For more information, we provide open project repositories for the \texttt{Astromer} project and its variants.\footnote{\url{https://github.com/astromer-science/multiband-astromer}} Documentation, tutorials, and data are provided within. Contributions via pull requests are welcome.

\begin{acknowledgements}

This work made use of data from the \texttt{MACHO} project. We acknowledge the support of the Institute of Geophysics and Planetary Physics, Lawrence Livermore National Laboratory, U.S. Department of Energy; the NSF-supported Center for Particle Astrophysics at the University of California at Berkeley, Santa Barbara, and San Diego; and the Australian National University.

This work also made use of data from the \texttt{ATLAS} project. The \texttt{ATLAS} project is primarily funded to search for Near-Earth asteroids through NASA grants NN12AR55G, 80NSSC18K0284, and 80NSSC18K1575; byproducts of the NEO search include images and catalogs from the survey area. This was partially funded by Kepler/K2 grant J1944/80NSSC19K0112 and HST GO-15889, and STFC grants ST/T000198/1 and ST/S006109/1. The \texttt{ATLAS} science products have been made possible through the contributions of the University of Hawaii Institute for Astronomy, the Queen's University Belfast, the Space Telescope Science Institute, the South African Astronomical Observatory, and The Millennium Institute of Astrophysics (MAS), Chile.

The authors acknowledge the contributions of the original \texttt{Astromer} development team and associated collaborators.

\end{acknowledgements}

\bibliographystyle{bibtex/aa}
\bibliography{bibtex/references}

\begin{appendix}

\section{Input transformation and attention mechanism of \texttt{Astromer}}\label{appendix:astromer}

\subsection{Input transformation}
Within the framework of \texttt{Astromer}, light curve analysis focuses on a single band. \texttt{Astromer} generates an embedding for each light curve by converting temporal and photometric information into a fixed-length vector. For a given light curve $ Z^{(k)} $, the embedding $Z^{(k)'}\in \mathbb{R}^{L \times 256} $ is used as the input for the multi-head attention blocks and is formulated as:
\begin{equation}
    Z^{(k)'} = \operatorname{PE}(t) + \mathbf{\hat{m}}^{(k)}\,W^\top,
\end{equation}
where $ \operatorname{PE}(t) \in \mathbb{R}^{L \times 256} $ represents the positional encoding of $ L $ observation times, $W^\top \in \mathbb{R}^{1 \times 256}$ is a learned weight matrix, and $\mathbf{\hat{m}} \in \mathbb{R}^{L \times 1}$ is a vector containing all observations
\begin{equation}
     \mathbf{\hat{m}}^{(k)} = \left\{  x_{0}^{(k)},   \dots, x_{j}^{(k)}, \ldots ,x_{m-1}^{(k)}\right\} \nonumber.
\end{equation}

To encode observation times, \texttt{Astromer} employs fixed sinusoidal positional encodings, following the approach introduced by \citet{vaswani2017attention}. This method uses trigonometric functions of varying frequencies to represent temporal positions. Although originally developed for evenly spaced sequences in natural language processing, this encoding is directly applied to the irregular sampling found in astronomical light curves as part of the original \texttt{Astromer} design.
\begin{equation}
    \operatorname{PE}_{j, t_l} =
    \begin{cases}
    \sin (t_l \times \omega_j), & \text{if } j \text{ is even}, \\[1mm]
    \cos (t_l \times \omega_j), & \text{if } j \text{ is odd},
    \end{cases}
\end{equation}
with the angular frequencies $\omega_j$ specified as
\begin{equation}
    \omega_j = \frac{1}{1000^{2j/d_{\operatorname{PE}}}},    
\end{equation}
where $ j \in [0,\dots,d_{\operatorname{PE}}-1] $ and $ d_{\operatorname{PE}} = 256 $ is the dimensionality of the positional encoding.

We note that the time encoding used in \texttt{Astromer} maps each scalar observation time to a high-dimensional embedding vector. As a result, this transformation does not correspond to a direct modification of the light curve in the original time--magnitude space, and therefore does not admit a simple ``before and after'' visualization analogous to a light curve plot. Instead, the role of the time encoding is to provide the transformer with a representation in which temporal relationships can be more easily learned in a higher-dimensional space.

\subsection{Self-attention mechanism}
Given an input sequence $Z$, each attention head first computes the queries, keys, and values as
\begin{equation}
    Q = Z W_Q^{\top}, \quad K = Z W_K^{\top}, \quad V = Z W_V^{\top},
\end{equation}
where $ W_Q$, $W_K$, and $W_V$ are the queries, keys, and values weight matrices, respectively. The masked attention output for pre-training is then derived via the scaled dot-product attention using a binary matrix $ M $ that has entries set to $ 1 $ when masked and $ 0 $ otherwise:
\begin{equation}
    \label{eq: softmax}
    z = \operatorname{softmax}\!\left( \frac{QK^\top + (-\infty)M}{\sqrt{d_h}} \right)V.
\end{equation}

In practice, the additive masking term $(-\infty)M$ used in Eq.~\ref{eq: softmax} is implemented as $\text{-}10^9M$ for numerical stability. This avoids missing values during softmax computation while effectively suppressing attention to masked positions. We follow this standard practice in our implementation, and note that the choice of constant does not significantly affect model behavior as long as it is sufficiently negative. The encoder ultimately produces a sequence of 200 contextual embeddings, each of length 256, which collectively capture the relationships between observations within a light curve.

\FloatBarrier 
\twocolumn

\end{appendix}

\end{document}